\documentclass[superscriptaddress,aps,prd]{revtex4-1}
\usepackage{amsfonts}
\usepackage{graphicx}
\usepackage{amsmath}
\usepackage{color}
\usepackage[english]{babel}
\usepackage[T1]{fontenc}

\usepackage{amsmath,amssymb}
\usepackage{url}
\usepackage{empheq}
\usepackage{fancyhdr}
\usepackage{epsfig}
\usepackage{caption}
\usepackage{subcaption}
\usepackage{url}

\newcommand{\mbf}{\mathbf}

\newcommand{\be}{\begin{equation}}

\newcommand{\ee}{\end{equation}}

\newcommand{\gv}[1]{\ensuremath{\mbox{\boldmath$ #1 $}}}

\begin{document}

\title{The magnetoelectric effect due to a semispherical capacitor surrounded by a spherical topologically insulating shell }

\author{Daniel G. Vel\'azquez}
\email{danielgv@ciencias.unam.mx}
\affiliation{Instituto de Ciencias Nucleares, Universidad Nacional Aut\'{o}noma de M\'{e}xico, 04510 M\'{e}xico, Distrito Federal, M\'{e}xico}

\author{L. F. Urrutia}
\email{urrutia@nucleares.unam.mx}
\affiliation{Instituto de Ciencias Nucleares, Universidad Nacional Aut\'{o}noma de M\'{e}%
xico, 04510 M\'{e}xico, Distrito Federal, M\'{e}xico}

\date{\today }

\begin{abstract}
We consider the magnetoelectric effect produced by a  capacitor  
formed by two semispherical  perfectly conducting plates subjected to  a potential difference  and surrounded by a spherical  shell of a topologically insulating material. The modified Maxwell equations are solved in terms of coupled electric and magnetic scalar potentials using spherical coordinates and in the approximation where the effective  magnetoelectric coupling is of the order of the fine structure constant. The emphasis is placed in the calculation of the magnetic field for several relevant   configurations designed to enhance the possibility of measuring  this field. The magnitudes we obtain  fall within the sensitivities
of magnetometers based upon NV centers in diamond as well as of devices using scanning SQUID magnetometry.   
\end{abstract}

\maketitle

\section{Introduction}
Topological insulators (TIs) are recently discovered materials which exhibit remarkable properties originating from  their peculiar  band structure. From an effective  macroscopic perspective, which corresponds to the approach we consider in this work, they can be characterized as  insulators in the bulk, but conductors on their  surfaces due to the presence of   quantized Hall currents. The recognition of topological phenomena in condensed matter dates back to Ref. \cite{twentytwo} which identified the conductivity of the quantum Hall effect \cite{fourteen} with the first Chern number of a Berry curvature in the reciprocal space.
Since then, several investigations \cite{sixteen}-\cite{twelve} led to a more profound understanding of these topological phases, both theoretically and experimentally. In particular, Bernevig \cite{one} predicted the existence of two-dimensional TIs in quantum wells of ${\rm{HgTe}}$, and soon König confirmed it experimentally \cite{fifteen}. {In subsequent years, the phenomenon became generalized to three dimensions, again starting with  theoretical predictions \cite{FUKANE,FUMELE,MB, monopole3,ROY} followed by experimental  confirmation \cite{ten}.}

The equations that encode the electromagnetic response of ordinary matter can be derived from the standard  Lagrangian density in electrodynamics $\mathcal{L}_{\rm em}=(1/8\pi)\Big(\left[\varepsilon\mathbf{E}^2-(1/\mu)\mathbf{B}^2 \right]+\mbf{J}\cdot \mbf{A}-\rho \Phi \Big)$, once the fields are expressed in terms of the electromagnetic potentials ${\mathbf A}, \Phi$. Here  $\varepsilon$ and $\mu$ are the  permittivity and the 
permeability of the medium while  $\rho$ and $ \mbf{J}$ stand for the external charge density and current density, respectively.

The electromagnetic response of TIs is captured by adding the term $\mathcal{L}_{\vartheta}=(\alpha /4 \pi^2)\vartheta(\mathbf{r},t)\mathbf{E}\cdot\mathbf{B}$ to the Lagrangian density   $\mathcal{L}_{em}$, where $\alpha=e^2/\hbar c$ is the fine structure constant and $\vartheta$  describes an additional property of the medium called  the magnetoelectric polarizability (MEP) \cite{0810}.  The major  modification that results from adding this term  
is the so-called {magnetoelectric effect} (MEE), which consists in the induction of a magnetization by an electric field and/or a polarization by a magnetic field \cite{fiebig}.

A simple way to show the properties  of $\mathcal{L}_{\vartheta}$ is to introduce the Faraday tensor $F_{\mu\nu}=\partial_{\mu}A_{\nu}-\partial_{\nu}A_{\mu}$ and the Levi-Civita symbol $\varepsilon^{\mu \nu \rho \sigma}$, in terms of which we  rewrite $\mathbf{E}\cdot \mathbf{B}= -\frac{1}{8} \varepsilon^{\mu \nu \rho \sigma}F_{\mu\nu }F_{\rho \sigma} \equiv -\frac{1}{8} {\cal P}$, where ${\cal P}$ is the abelian  Pontryagin density, which is a topological invariant \cite{TINV}.  The Bianchi identity yields 
$\mathbf{E}\cdot \mathbf{B}=-\frac{1}{4} \varepsilon^{\mu \nu \rho \sigma}\partial_\mu(A_{\nu}F_{\rho \sigma})$, which implies that when 
$\vartheta$ is constant, the term $\mathcal{L}_{\vartheta}$ is a total derivative, so it does not affect the equations of motion. 

The topological properties induced by the coupling to the MEP are most clearly appreciated  by defining the action of the system as 
$S=\frac{\hbar}{e^2 \alpha} \int d^4 x ({\cal L}_{\rm em}+ {\cal L}_\vartheta)$ in the CGS system, where the electric and the magnetic fields  have the same units of charge divided by square distance. 
Assuming periodic boundary conditions and a manifold without boundaries, the contribution $S_\vartheta$ results 
\begin{equation}
\frac{S_\vartheta}{\hbar}=\frac{\vartheta}{32 \pi^2}\int d^4x \epsilon^{\alpha\beta\mu\nu} \frac{1}{e^2} F_{\alpha\beta} F_{\mu\nu} \equiv \vartheta C_2,
\end{equation}
where the integer $C_2$ is the second Chern number of the manifold. Recalling that the relevant physical object is
$e^{i S_\vartheta/\hbar}$ we realize that the extended electrodynamics 
is invariant under  the transformation $\vartheta\rightarrow \vartheta +2\pi n$ \cite{wilk,MVMF}. Moreover, imposing time reversal invariance $\vartheta \rightarrow - \vartheta$ yields 
the two possible values $\vartheta=0$ and $\vartheta=\pi$ (modulo $2\pi$), which  satisfy $e^{i \vartheta}=e^{-i \vartheta}$  \cite{0810}. This produces the $\mathbb{Z} _{2}$ classification of insulators where $\vartheta=0$ characterizes normal insulators, whereas $\vartheta=\pi$ defines the topological phase. To detect the MEE in TIs we need at least an interface between two media with different  values $\vartheta_1$ and $\vartheta_2$ such that $\partial_\mu \vartheta \neq 0$ there. A smooth transition between these values requires to break time reversal symmetry, which is usually achieved by coating the interface with a thin magnetic layer a few nanometers wide. Since the values of $\vartheta$  are determined modulo $2\pi$, we will have  $\vartheta_1 -\vartheta_2=\pi +2\pi n$ with  $n$ to be determined by the specific time reversal symmetry breaking  mechanism  at the interface.   The non-trivial coefficient $ (2n+1)\pi$ is related to a semi-quantized conductivity on the surface of the material $\sigma=\frac{(\vartheta_1-\vartheta_2) e^2}{2\pi h}= \frac{(n+1/2)e^2}{h}$ which arises due to the boundary terms in ${\cal L}_\vartheta$  \cite{monopole}. Here $h= 2 \pi \hbar$ is the Planck constant. The physical origin of this phenomenon in topological materials is the quantum Hall effect \cite{3}. In general, the conductivity acquires the form $\sigma=\frac{e^2}{h}\nu$, where $\nu$ can take very specific integer and rational values. The last of these values  is attained in the fractional QHE due to fractionally charged quasiparticles (anyons) that are neither bosons nor fermions \cite{tong}. 

An interesting aspect of the MEE  in TIs is that an electric charge near its surface can generate not only an image electric charge as usual, but also a magnetic image monopole, which provides an alternative mathematical  interpretation of the  magnetic field produced by the  surface currents resulting from the quantum Hall effect [\citealp{monopole,3}]. 

Let us  observe that the equations which arise from the Lagrangian density $\mathcal{L}_{\rm em} + \mathcal{L}_{\vartheta}$ may describe different physical phenomena according to the choice of $\vartheta(t, \mathbf{x})$, and not only the electromagnetic response of TIs.  For example, the electrodynamics of metamaterials when $\vartheta  \, \in \,  \mathbb{C}$ [\citealp{magnetoelectric,magnetoelectric12}] and the response of Weyl semimetals  when $\vartheta(\mathbf{x},t)=2\mathbf{b}\cdot \mathbf{x}-2b_0t$ [\citealp{magnetoelectric,magnetoelectric13,4}]. The term $\mathcal{L}_{\vartheta}$  also describes the interaction of the hypothetical axionic field with the electromagnetic field in elementary particle physics [\citealp{3,33}].  In the last decade, the MEE was also reproduced in composite structures \cite{ferrites}, which opened the door to studies with a view to the manufacture of new devices. The first composite magnetoelectric material was created from the ferroelectric BaTiO$_3$ and the ferromagnetic CoFe$_2$O$_4$ \cite{fiebig}. These composite materials have shown more intense couplings than the monophasic ones which exhibit polarization and magnetization in the same phase.

To obtain  the modified Maxwell equations  from the Lagrangian density ${\cal L}_{em}+{\cal L}_{\vartheta}$ we  introduce the electromagnetic potentials $\Phi, {\mathbf A}$ such that 
\begin{equation}
{\mathbf E}=-\frac{1}{c} \frac{\partial {\mbf A}}{\partial t}-\nabla \Phi,\quad {\mbf B}=\nabla \times  {\mbf A},
\end{equation} which leads to the standard homogeneous Maxwell equations $\nabla \cdot {\mbf B}=0$ and  $c \nabla \times  {\mbf E}=- {\partial {\mbf B} }/{\partial t}$.  
The resulting inhomogeneous  equations are
\begin{equation}
\nabla \cdot (\epsilon \mathbf{E})=4\pi \rho +\frac{\alpha}{\pi} \nabla \vartheta \cdot \mathbf{B}, \quad
\nabla \times (\mathbf{B}/ \mu)-\frac{1}{c}\frac{\partial (\epsilon \mathbf{E})}{\partial t}=\frac{4\pi}{c}\mathbf{J}-\frac{\alpha}{\pi}\nabla \vartheta \times \mathbf{E}-\frac{1}{c}\frac{\alpha}{\pi}\frac{\partial \vartheta}{\partial t}\mathbf{B}.
\label{INHEQ}
\end{equation}
They can also  be understood  as  those of electrodynamics in a material medium having  the  constitutive relations
\begin{equation}
\mathbf{D}=\epsilon\mathbf{E}-\frac{\vartheta \alpha}{\pi}\mathbf{B}, \quad \mathbf{H}=\frac{1}{\mu}\mathbf{B}+\frac{\vartheta \alpha}{\pi}\mathbf{E}.
\label{consteq}
\end{equation}
When $\vartheta({\mbf x})$ takes constant values,   $\vartheta_1$  and $\vartheta_2, $ in two regions ${\cal U}_1$ and  ${\cal U}_2$ separated by an interface $\Sigma$ parametrized with the equation $F_\Sigma({\mbf x})=0$,
the  effective sources arising from Eqs. (\ref{INHEQ}) are 
\begin{equation}
\rho_\vartheta= \frac{\alpha}{4 \pi^2}(\vartheta_2 -\vartheta_1)\delta(F_\Sigma({\mbf x})){\mbf n}_\Sigma \cdot {\mbf B}, \qquad {\mbf J}_\vartheta=\frac{c \alpha}{4 \pi^2}(\vartheta_2 -\vartheta_1)\delta(F_\Sigma({\mbf x})){\mbf n}_\Sigma \times {\mbf E}, 
\end{equation} 
where ${\mbf n}_\Sigma$ is a vector perpendicular to the interface. That is to say, corrections to the dynamics arise only at the interface, while the bulk regions satisfy the unmodified Maxwell equations. A particularly interesting case occurs when $\Sigma$ coincides with the surface of  a perfect conductor, since then $\mbf E$ is normal to the interface, while ${\mbf B}$ is tangential, yielding that both effective sources are zero, in spite of the presence  of a gradient of $\vartheta({\mbf x})$. It is important to note that this result is general: if a $\vartheta$ interface matches a perfect conducting  surface, there is no contribution from that interface to the Maxwell equations (\ref{INHEQ}).

The paper is organized as follows. In Section \ref{statement} we establish our general setup and adapt the general equations (\ref{INHEQ}) 
to this case. The boundary conditions at the interfaces are also written, allowing us  to match the solutions to the standard Maxwell equations in each bulk region.  Since the general solution   depends on ${\tilde \alpha}\equiv (\vartheta_2-\vartheta_1)\alpha/\pi$ in a way that is hard to handle, we use the fact that $\alpha$  is of order $10^{-2}$, and that $\vartheta_{1,2}$ are of order unity, to consider a perturbative expansion in $\tilde \alpha $. The detailed procedure is found in the Appendix. In Section \ref{solution}, we consider some particular configurations (limiting cases) to simplify the problem on one side and, on the other, to confer more physical relevance to the results. Also  we show several plots of the streamlines of the fields $\mathbf{E}$ and $\mathbf{B}$ for some of these configurations that intend  to be representative of the limiting cases above mentioned.  In Section \ref{experimental}, we show some estimations of the magnetic field produced by  the MEE in this hemispheric system, which turn out to be detectable within the current experimental possibilities. Nowadays, it is possible to measure magnetic fields of order $10-100$ mG \cite{revtex36,revtex37} and magnetic fluxes of the order of $10^{-14}$ Gcm$^2$ \cite{SQUID}. This bestows phenomenological relevance to the problem. Our results suggest two different empirical approaches: of the two most relevant configurations, one generates highly isotropic but weak fields (though detectable), while the other gives rise to intense fields in particular directions.
\section{The magnetoelectric effect in a  semiespherical capacitor surrounded by a spherical shell made of a TI }
\label{statement}
One of the simplest way to explore the MEE in topological materials is to locate an electric source in front of them and determine/measure the induced magnetic field. Implementation of this idea has been reported for the following cases: (i)  a pointlike charge in front of a planar TI \cite{monopole}, (ii) a sphere of finite radius in front of a planar TI \cite{esfera} and (iii) a pointlike charge in front of a Weyl semimetal \cite{PRB}.

\begin{figure}[htb!]
\includegraphics[width=0.4\linewidth]{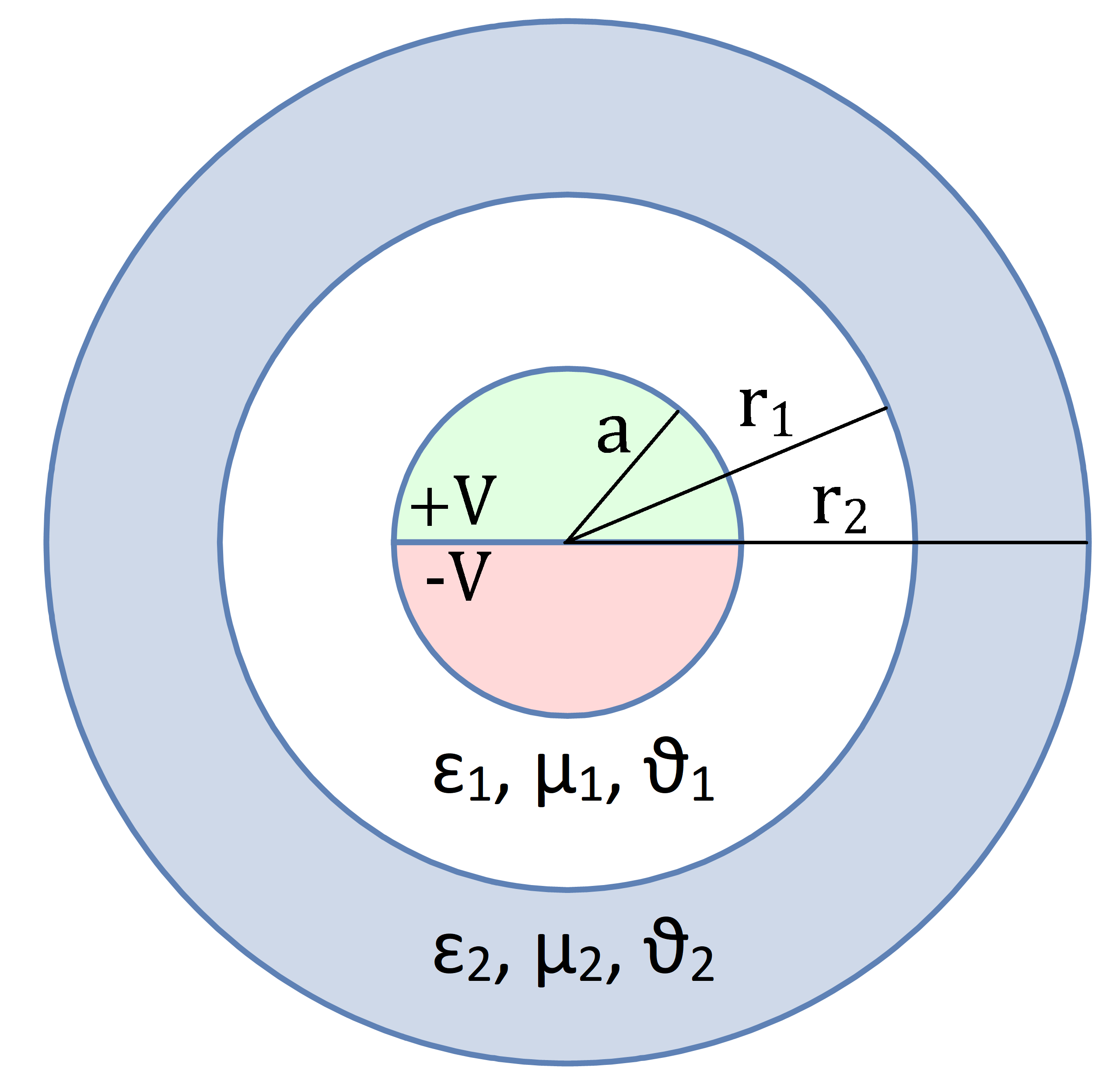}
\caption{System composed by a  semiespherical capacitor of radius $a$ surrounded by a thick TI shell of width $r_2-r_1$. }
\label{arreglo}
\end{figure}

The problem we consider here is that of a  capacitor  
formed by two semispherical plates of radius $a$ that are kept at different potentials ($V$ and $-V$). We call this a semiespherical capacitor. The region 1 ($ a < r <r_1$) is filled with a media with permittivity $\varepsilon_1$ and MEP $\vartheta_1$. At a distance $r_1$ from its center,  such that $r_1 \geq a$, there is a thick TI shell of width $r_2-r_1$ defining the region 2 ($r_1  <r < r_2$) and having permittivity $\varepsilon_2$ and MEP $\vartheta_2$. The region 3 ($r_2 < r < \infty$)  has the same parameters as the region 1. We take  regions 1 and 3 to be the vacuum, such that  $\varepsilon_1=1, \, \vartheta_1=0$, and we consider non-magnetic materials taking $\mu=1$ everywhere. The setup is shown in Fig. \ref{arreglo}.
In the case that concerns us, the MEP $\vartheta ({\mathbf x})$  displays spherical symmetry and is given by
\begin{equation}
\vartheta(r)= \vartheta_1 H(r-a)H(r_1-r)+ \vartheta_2 H(r-r_1) H(r_2-r)+\vartheta_1 H(r-r_2),
\end{equation}
where $H(r)$ is the Heaviside function.
The two inhomogeneous Maxwell equations in Eq. (\ref{INHEQ}) become
\begin{equation}
\nabla \cdot(\epsilon  \mathbf{E})=\tilde{ \alpha}\Big( \delta(r-r_1)-\delta(r-r_2)\Big)\mathbf{B}\cdot \mathbf{\hat{r}}, \label{INHOM1} 
\end{equation}
\begin{equation}
\nabla \times \left(\frac{ \mathbf{B}}{\mu}\right) -\frac{1}{c}\frac{\partial (\epsilon {\mathbf E})}{\partial t}=\tilde{ \alpha} \Big( \delta(r-r_1)-\delta(r-r_2)\Big)\mathbf{E}\times \mathbf{\hat{r}},
\label{INHOM2} 
\end{equation}
where we recall  the parameter 
\begin{equation}
\tilde{ \alpha}=(\vartheta_2-\vartheta_1)\alpha/\pi, 
\end{equation}
which summarizes the whole topological effects.
As explained before, the contribution at $\delta(r-a)$ does not appear in the above equations because the capacitor plates are perfect conductors.
Here $ \mathbf{\hat{r}}$ is the unit vector in the radial direction. Since the MEP's are constant in each bulk region, the dynamical modifications only arise at the spherical  interfaces $\Sigma_{1}, \,  \Sigma_2$ located at $r=r_1, \, r=r_2$, respectively, where a discontinuity in the MEP arises. They are coded in the boundary conditions at the interfaces, which are obtained from Eqs. (\ref{INHOM1}) and (\ref{INHOM2}). Alternatively, using the constitutive relations (\ref{consteq}),  one can determine them by imposing at the interfaces the continuity of the normal components of $\mbf{D}$ and  $\mbf{B}$ together with the continuity of the parallel components of $\mbf{H}$ and  $\mbf{E}$, in the absence of external sources.  The results at the interface $\Sigma_1$ located at $r=r_1$ are 
\begin{eqnarray}
&&\Big[\varepsilon\mathbf{E}\cdot \mbf{\hat{r}}\Big]_{r=r_1}= \tilde{ \alpha}(\mathbf{B}\cdot \mbf{\hat{r}})_{r=r_1}, \qquad \,\,\, \,\,\,\, \Big[\mathbf{B} \times \mbf{\hat{r}}\Big]_{r=r_1} = -\tilde{ \alpha}(\mathbf{E}\times\mbf{\hat{r}})_{r=r_1}, 
 \nonumber \\
&&\Big[\mathbf{B}\cdot\hat{{\mbf r}}\Big]_{r=r_1}=0, \qquad \qquad \qquad  \qquad  \Big[\mathbf{E} \times \hat{{\mbf r}}\Big]_{r=r_1}=0,\label{cb}
\end{eqnarray}
The results at the interface  $\Sigma_ 2$ at $r=r_2$ are obtained from the above just making ${\tilde \alpha} \rightarrow -{\tilde \alpha}$
Here $\Big[Q\Big]_{r=b}\equiv \lim_{\delta \rightarrow 0}(Q(r=b+\delta)-Q(r=b-\delta))$, denotes  the discontinuity of $Q$ at the spherical interface located at $r=b$, while $(S)_{r=b}\equiv S(r=b)$ is the evaluation of a continuous function $S$ at the corresponding interface.

The MEE is physically realized by the generation of surface currents $\mbf K$  at the interfaces, due to the magnetization $\mbf M=-\frac{\tilde \alpha}{4 \pi} {\mbf E}$, such that 
\begin{equation}
{\mbf K}_{\vartheta, I}=-\frac{\tilde \alpha}{4 \pi} {\mbf E}\times {\mbf {\hat n}}|_{r=r_I}, \label{SURCUR}
\end{equation} 
where $I=1,2$ denotes the corresponding interfaces $\Sigma_1$ and $\Sigma_2$ respectively. Here $\mbf {\hat n}$ in the unit normal exterior to the surface of the TI. These currents are the physical sources of   the magnetic field.

For the resolution of the equations  in the bulk  we use scalar electric and magnetic potentials $\Phi$ and $\Psi$, a choice that is allowed by the conditions $\nabla \times \mathbf{E}=0$ and $\nabla \times \mathbf{B}=0$ in each bulk, such that
 \begin{equation}
 \mathbf{E}=-\nabla \Phi, \qquad \qquad \mathbf{B}=-\nabla \Psi.
 \end{equation}
Since the electric and magnetic fields satisfy the homogeneous Maxwell equations $\nabla \cdot \mathbf{E}=0$ and $\nabla \cdot \mathbf{B}=0$, the aforementioned potentials satisfy Laplace equation in  the  regions 1, 2 and 3 previously defined.
As the system possesses azimuthal symmetry with respect to the axis ($z$-axis) perpendicular to the plane ($x-y$ plane) that separates de two semispherical plates, it is enough to express the general solution in spherical coordinates ($x=r\sin \theta \cos \phi, \,\, y=r\sin \theta \sin \phi, \,\ z=r\cos \theta$) as
\begin{eqnarray} \label{gen_elec}
\Phi_i (r,\theta)&=&\sum_{l=0}^\infty(A_l^i r^l+B_l^i r^{-(l+1)}) P_l(\cos \theta),\\ \label{gen_mag}
\Psi_i (r,\theta)&=&\sum_{l=0}^\infty(C_l^i r^l+D_l^i r^{-(l+1)}) P_l(\cos \theta),
\end{eqnarray}
where the index $i$ can take the values 1, 2, and 3, referring to the regions defined  before.
This reduces the whole inquiry to the determination of the  coefficients $A_l^i, B_l^i, C_l^i, D_l^i$  according to the boundary conditions in Eq. (\ref{cb}), together with the similar ones at  $r=r_2$. The Appendix contains a  detailed discussion of the solution for the potentials $\Phi_i$  and $\Psi_i$. Let us emphasize that the boundary condition $V(\theta)=-V(-\theta)$ on the surface of the capacitor implies that all the coefficients with even values of $l$ are zero. There are twelve coefficients for every odd value of $l$ and twelve equations that relate them linearly. From now on we introduce the notation 
\begin{equation}
\sum_{l}{}^{\prime}\equiv \sum_{l=1,3,...}^\infty.
\end{equation}
Let us emphasize that our setup has a discontinuity of the electric potential at the equator of the capacitor, i.e. at $r=a, \, \theta=\pi/2$, so that approaching to these particular  points  in any calculation has to proceed as a limiting process. Since the electric potential is the source of the MEE, we expect that similar care is required when dealing with the magnetic field at these points.

\section{The Solution in power series of $\tilde \alpha$}
\label{solution}
\begin{figure}[htb!]
\centering
\begin{subfigure}{0.5\linewidth}
  \centering
\includegraphics[width=0.8\linewidth]{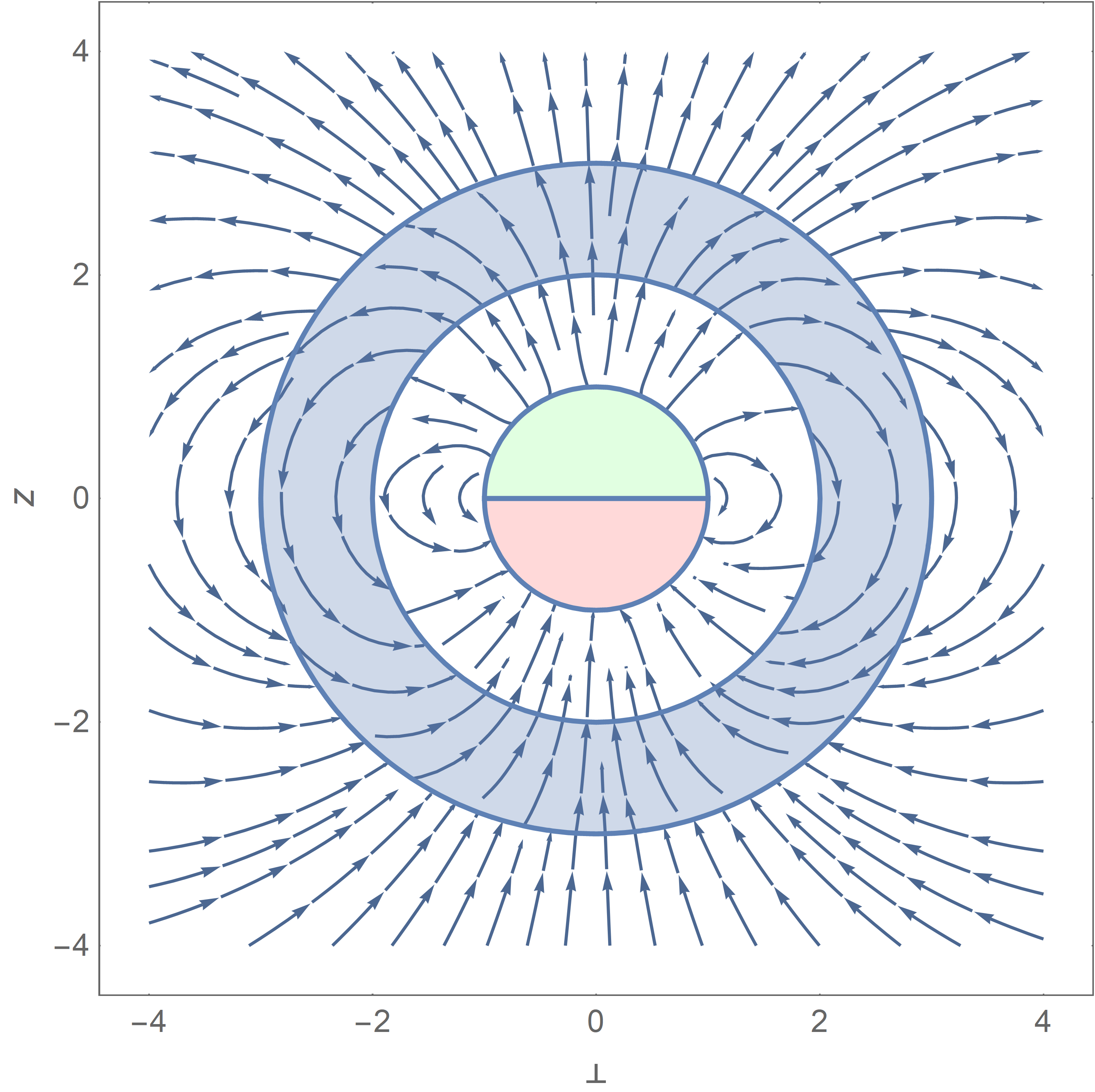}
  \caption{}
\end{subfigure}%
\begin{subfigure}{0.5\linewidth}
  \centering
\includegraphics[width=0.8\linewidth]{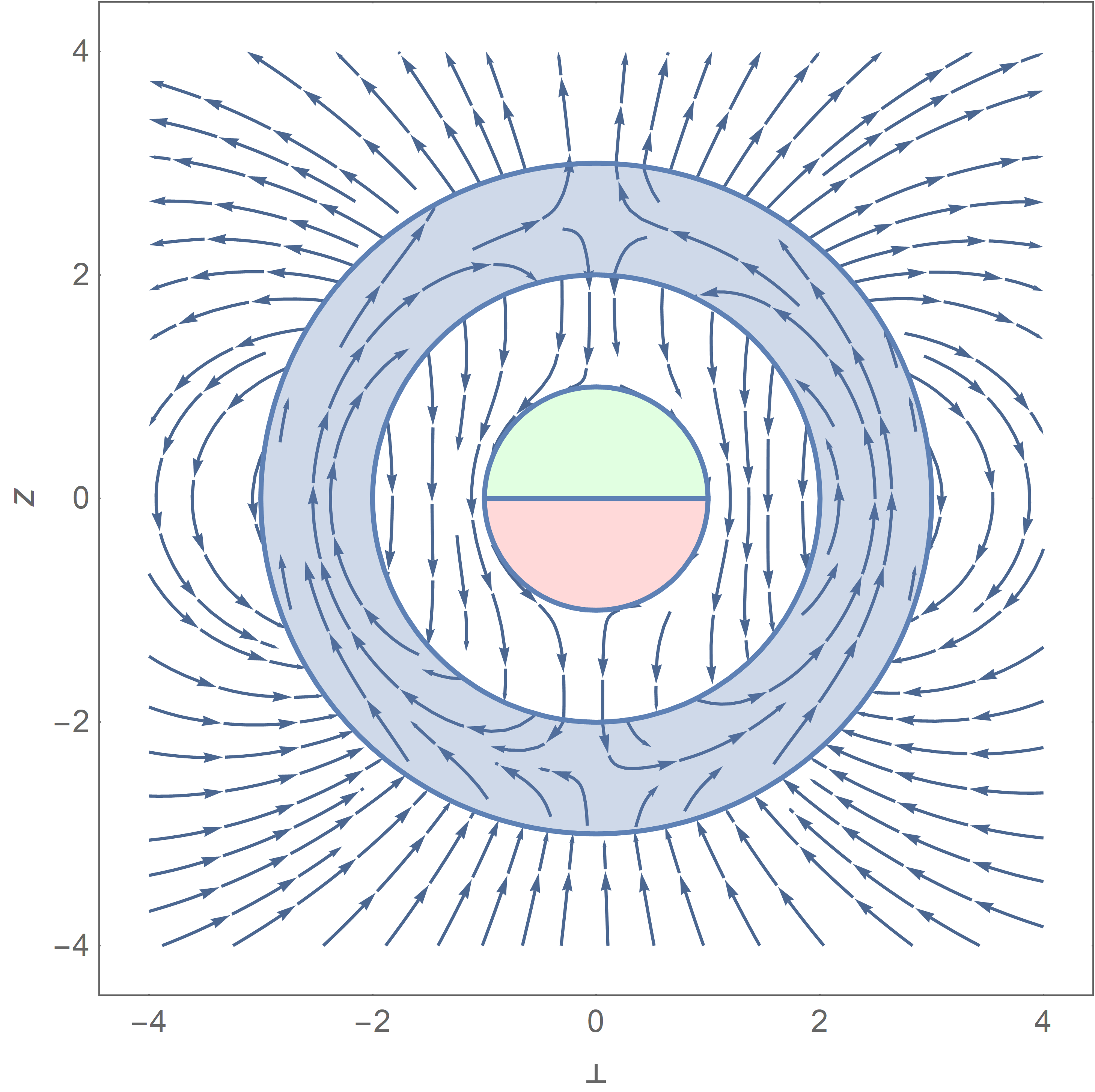}
  \caption{}
\end{subfigure}
\caption{Streamlines for a representative of the general case with  $a=1 \, \mu$m, $r_1=2 \ \mu$m, $r_2=3 \, \mu$m and $\varepsilon=4$. Panel (a) the electric field. Panel  (b) the magnetic field.}
\label{general}
\end{figure}

The full system of equations (\ref{SE}-\ref{MF}) for the coefficients is very involved and their solution is not to much illuminating. We will consider an approximate solution motivated by time reversal invariance (TRI) and the  fact that 
$\tilde \alpha$ is of the order of the fine structure constant. TRI 
imposes some restrictions on the $\tilde \alpha$ dependence of the coefficients according to whether their origin is electric or magnetic. In fact,  under time reversal symmetry ($\mathcal{T}$) the electric and magnetic fields transform according to $\mathcal{T} \mbf{E} = \mbf{E}$ and $\mathcal{T} \mbf{B} = - \mbf{B}$, while the charge and current densities behave as $\mathcal{T} \rho =  \rho$ and $\mathcal{T} \mbf{J} = - \mbf{J}$. Also $\mathcal{T}$ takes $\tilde{\alpha}$ into $- \tilde{\alpha}$ in agreement with  the $\mathbb{Z} _{2}$ classification of time-reversal invariant TIs. In our case, both the electric and magnetic fields are sourced by charge densities which create  the potential on the capacitor plates. Then, the linear relation between sources and fields, here constructed as gradients of the potentials, demand that $\Phi(\tilde \alpha)=\Phi(-\tilde \alpha)$  while $\Psi(\tilde \alpha)=-\Psi(-\tilde \alpha)$  to guarantee the correct behavior  of the corresponding fields under TRI \cite{esfera}. Our strategy is to look for an expansion of the coefficients in powers of $\tilde \alpha$, which we expect to converge rapidly due to the smallness of the  expansion parameter. Then, the 
previous reasoning requires that the electric (magnetic)   coefficients 
 $A^i_l, B^i_l$ ( $C^i_l, D^i_l$) include  only even (odd)  powers of  $\tilde \alpha$. We keep
 the approximation to the lowest order in $\tilde \alpha$, retaining only  the first order in the magnetic contributions, together with the zeroth order in the electric potential.   
Even in this simplified situation, the solutions for $A_l^i$, $B_l^i$, $C_l^i$ and $D_l^i$ in Eqs. (\ref{A51}-\ref{A62}) are not easy to handle, so it is convenient to discuss some particular configurations of the general setup in Fig. \ref{arreglo} to obtain more accessible results from the theoretical point of view, which in turn will allow  to interpret their physical consequences more easily. In the subsequent examples we take medium 1 and medium 3 as the vacuum ($\varepsilon_1=\varepsilon_3=1, \vartheta_1=\vartheta_3=0$),  and medium 2 as a TI with $\varepsilon_2=4$ and $\vartheta_2=\pi$. 

Before dealing with some particular cases we plot the streamlines of the fields for the configuration $a=1$, $ r_1=2 $, $r_2=3$ in  Fig. \ref{general},  which is intended to represent the general case. The shape of the electric field is explained by the fact that the permittivity of the magnetoelectric medium has been taken as distinct from that of the vacuum. In contrast, the magnetic field shows a more variable behavior, and it is worth noting that within the medium it is almost completely tangential, except for the vicinity of $\theta=0 $ and $\theta=\pi $. { Notice that in spite of the discontinuity of the source at $r=a, \, \theta=\pi/2$, the fields are well behaved everywhere outside the capacitor.}

\subsection{Case $\varepsilon=1$, $\tilde{ \alpha}=0$}
\label{case1}
\begin{figure}[htb!]
\includegraphics[scale=0.6]{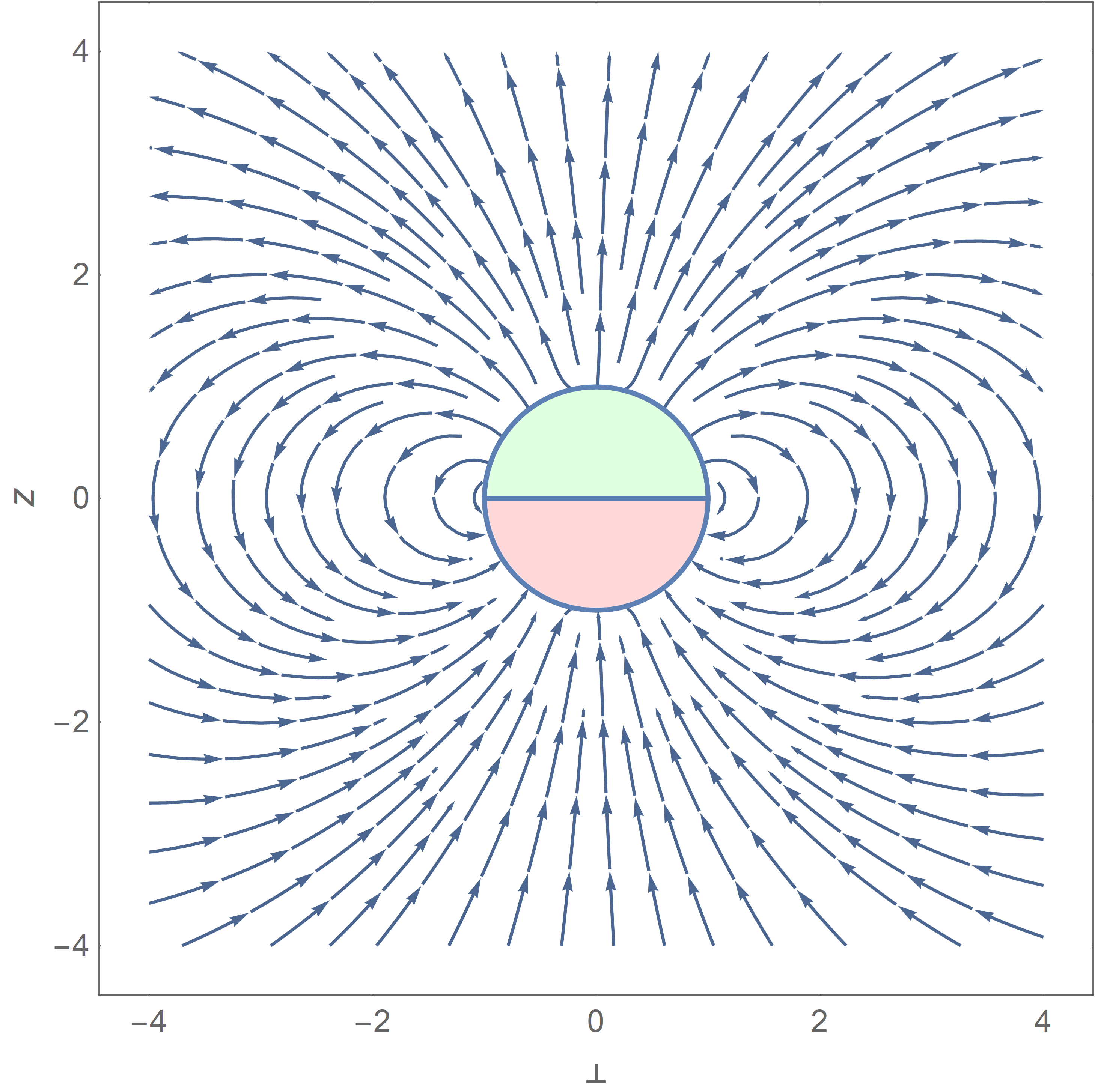}
\caption{Streamlines of the electric field for the trivial case.}
\label{electric_jackson}
\end{figure}

This first case consists in replacing  region 2, between $r_1$ and $r_2$, by the vacuum. This should imply null magnetic coefficients, since the magnetoelectric effect cannot manifest with a zero MEP. This problem corresponds to that of finding the electric field produced by a sphere with a northern hemisphere at potential $+V$ and a southern hemisphere at potential $-V$, which is solved in many textbooks (e.g. \cite{jackson}), and that we will call {the trivial configuration} henceforth. Under these conditions we verify that the system of equations (\ref{A51}-\ref{A62}) yields the only non-zero coefficients  
\begin{equation}
B_l^1=B_l^2=B_l^3=a^{l+1}V_l,
\end{equation}
with $V_l$ given by Eq. (\ref{V_l}).
Therefore, the magnetic field is identically zero throughout the space  due to the absence of MEP gradients, while the electric potential 
matches the results found in the literature \cite{jackson}
\begin{equation}
\Phi(r,\theta)=\sum_{l}{}^{\prime} \left(\frac{a}{r}\right)^{l+1} V_l P_l(\cos \theta). \label{TCONF}
\end{equation}
A plot of the streamlines of the electric field produced by the trivial configuration ($\varepsilon=1$, $\tilde{ \alpha}=0$, $r_1=r_2$, $r_1=a$, $r_2\rightarrow \infty$) is shown in Fig. \ref{electric_jackson}.
It is interesting to observe that the result  (\ref{TCONF}) is independent of $\varepsilon$ since the boundary conditions at the capacitor are stated in terms of the potential instead of the charges at the plates.
\subsection{Case $r_1=r_2$}
\label{case2}
This situation corresponds to the concurrence of the interfaces of the TI, which is now absent. Again we should expect to obtain zero  magnetic field. To verify this we focus on the induced currents $\mathbf{K}_{\vartheta ,I}$. In the limit we have 
\begin{equation}
\mathbf{K}_{\vartheta ,1}+\mathbf{K}_{\vartheta ,2}= -\frac{\tilde \alpha}{4 \pi} \lim_{r_1 \rightarrow r_2} \Big(\mathbf{E}\times \mbf{\hat{r}}|_{r=r1}-\mathbf{E}\times \mbf{\hat{r}}|_{r=r2} \Big)=0.
\end{equation}
 due to the continuity of the tangential component of the electric field. That is, the surface currents are canceled and no magnetic field is produced so that $\Psi=0$.  Therefore, the solution is reduced to the trivial case.

\subsection{Case $r_1=a$}
\label{case3}

This corresponds to eliminate the vacuum region between the capacitor and the 
TI, such that the interior of the shell touches the plates of the capacitor. Here the region $1$ shrinks to the surface of a sphere.
In order to consistently eliminate the region 1 we have to verify that the correct boundary conditions on the surface of the capacitor are now satisfied in terms of the coefficients $A_l^2, B_l^2, C_l^2, D_l^2 $ when $\Phi_2$ and $\Psi_2$ are evaluated at $r=a$. We establish these conditions after restricting  the coefficients  in Eqs. (\ref{A51}-\ref{A60}) to $r_1=a$, which yields  
\begin{eqnarray}
A_l^2&=&(l+1) (\varepsilon -1) {\lambda}^l_c, \qquad 
B_l^2= r_2^{2 l+1} (l \varepsilon +l+1){\lambda}^l_c,\qquad
B_l^3=(2 l+1) \varepsilon r_2^{2 l+1}{\lambda}^l_c, \nonumber\\
C_l^2&=&-\tilde{ \alpha}  (l+1) \varepsilon {\lambda}^l_c,\qquad 
D_l^2= -\tilde{ \alpha}  l \varepsilon  a^{2 l+1}{\lambda}^l_c,\qquad
D_l^3=-\tilde{ \alpha}  l \varepsilon  \left(a^{2 l+1}-r_2^{2 l+1}\right){\lambda}^l_c, \nonumber \\
&& \hspace{3cm}  {\lambda}^l_c=\frac{a^{l+1} V_l}{a^{2 l+1}(l+1) (\varepsilon -1) +r_2^{2 l+1} (l \varepsilon +l+1)} \label{COEFF_CASE_C}
\end{eqnarray}
The previous condition (\ref{BCPOT}) for the potential in the surface of the capacitor has to replaced by requiring $A_l^2 a^l +B_l^2 a^{-(l+1)}=V_l$ in terms of the coefficients of the region 2. From the results in Eq. (\ref{COEFF_CASE_C}) we verify that this relation is indeed satisfied. The  boundary condition (\ref{A34}) for the magnetic field at the capacitor should now reads $l a^{l-1}  C_l^2  = (l+1) a^{-(l+2)}  D_l^2$. Again, the values in Eq. (\ref{COEFF_CASE_C}) show that this condition is satisfied.  In this way we can safely forget about the region 1. In general, for this configuration the magnetic coefficients are non-zero in the two regions of interest. 

\begin{figure}[htb!]
\centering
\begin{subfigure}{.5\linewidth}
  \centering 
 \includegraphics[width=0.8\linewidth]{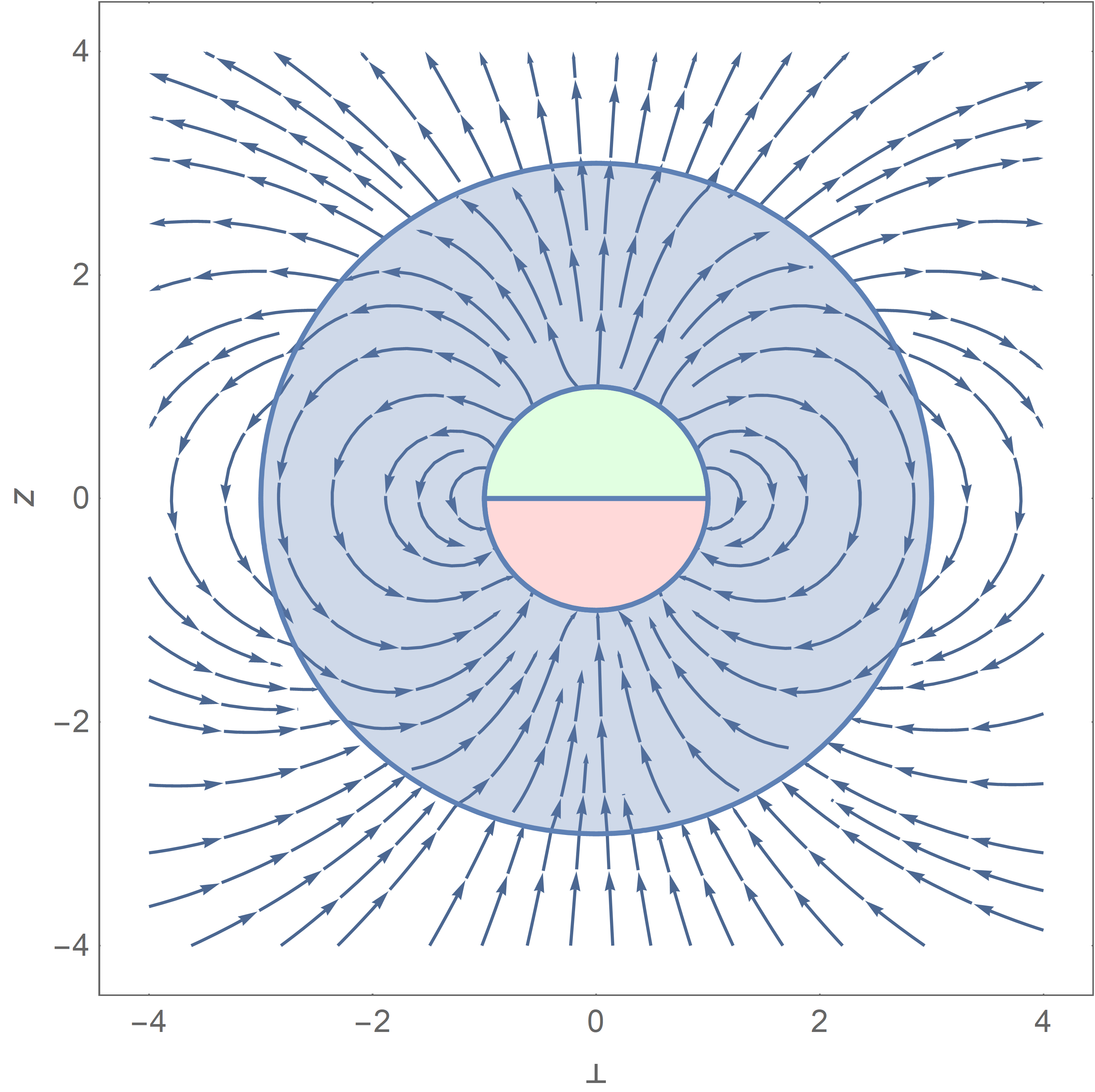}
  \caption{}
\end{subfigure}%
\begin{subfigure}{.5\linewidth}
  \centering
\includegraphics[width=0.8\linewidth]{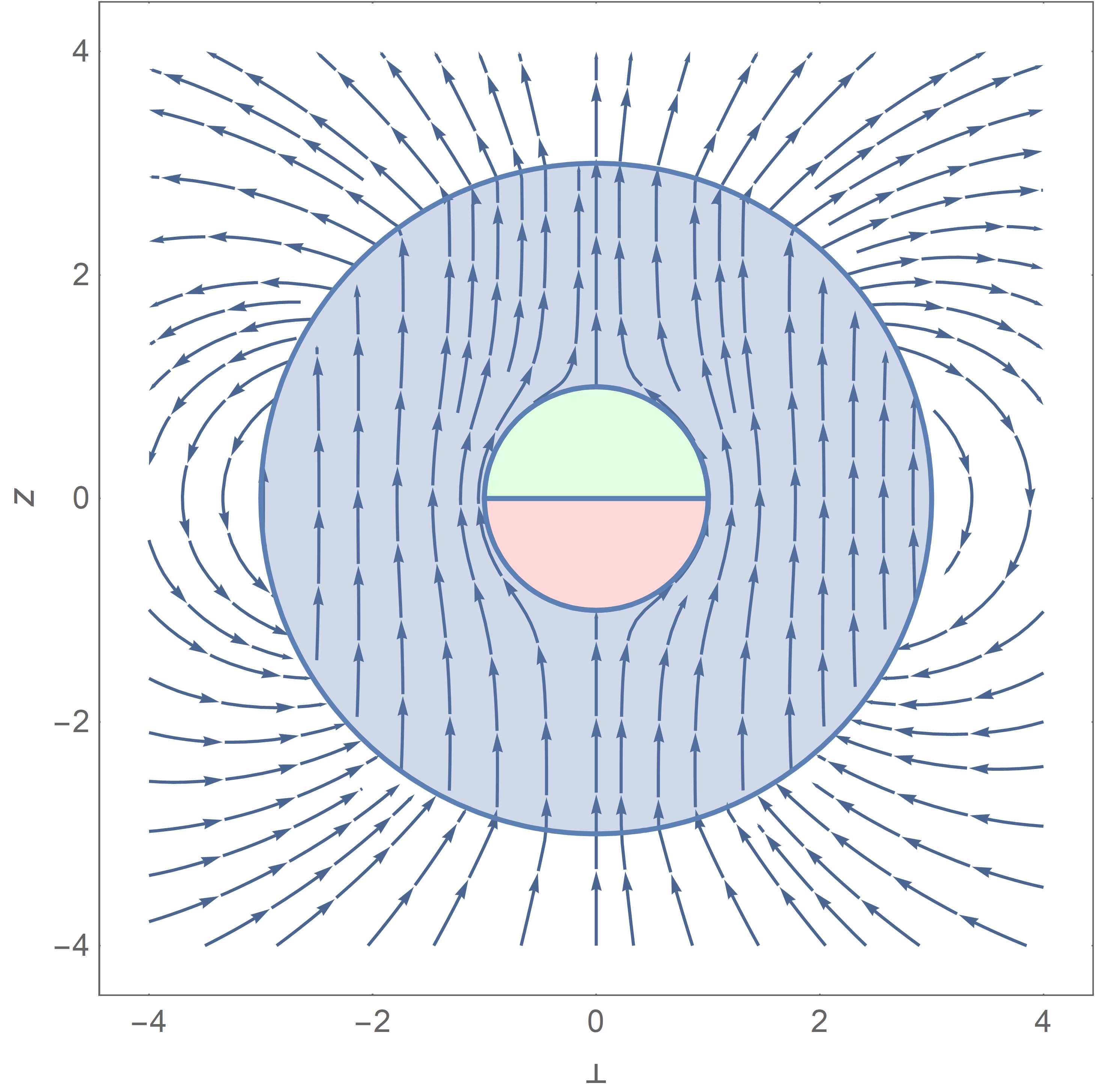}
  \caption{}
\end{subfigure}
\caption{Streamlines for  $a=r_1=1 \, \mu$m, $r_2=3 \, \mu$m and $\varepsilon=4$. Panel (a): the electric field  and Panel (b): the magnetic field. }
\label{r1=a}
\end{figure}

The case $a=r_1$ is plotted  in Fig. \ref{r1=a} which, as will be seen in Section \ref{experimental}, is of particular importance because it maximizes the field in the z-axis direction for a large range of values for  $a$. 

\subsection{Case $r_2\rightarrow \infty$}
\label{cas4}
\begin{figure}[htb!]
\centering
\begin{subfigure}{0.5\linewidth}
  \centering
  \includegraphics[width=0.8\linewidth]{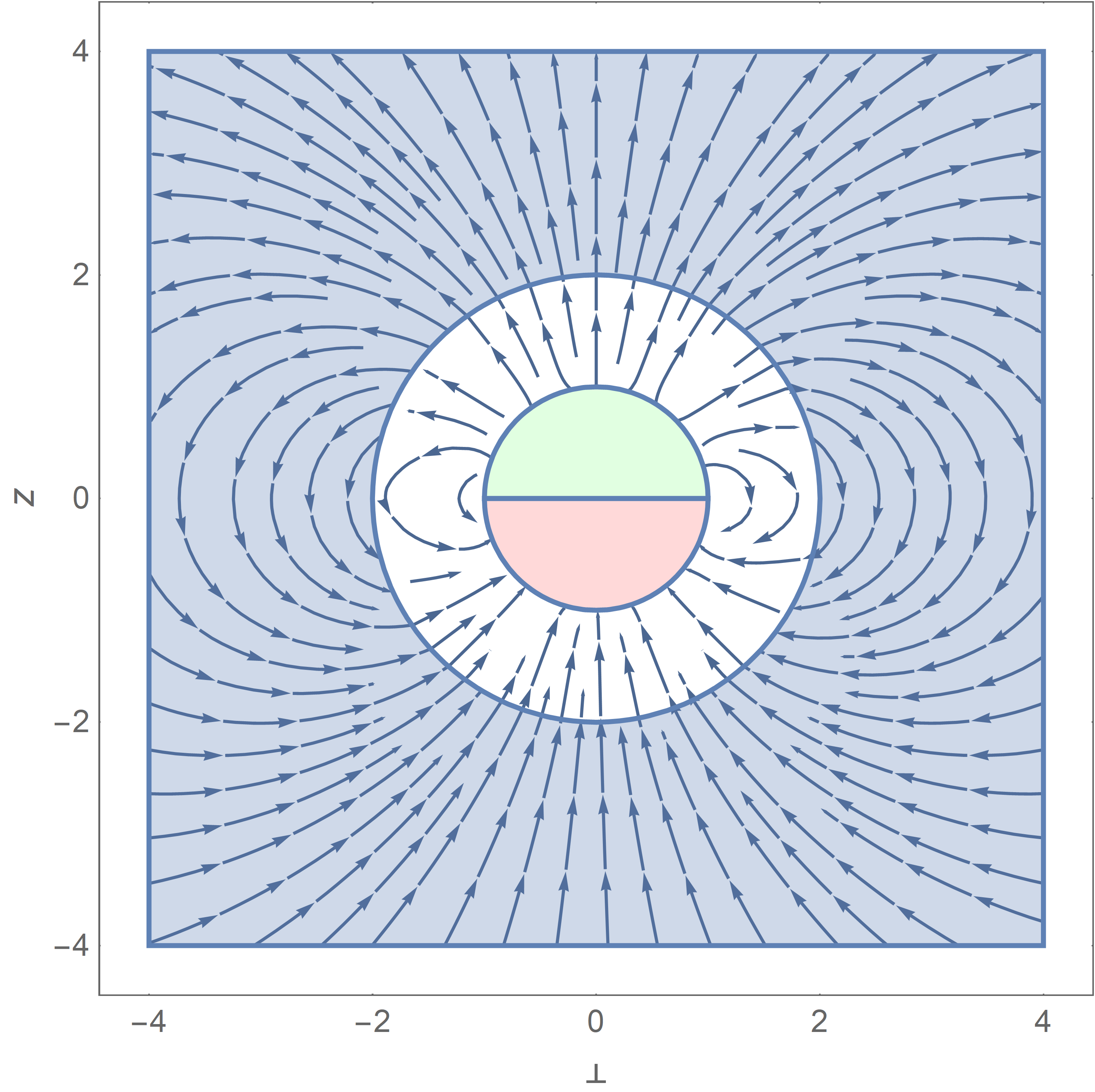}
  \caption{}
\end{subfigure}%
\begin{subfigure}{0.5\linewidth}
  \centering
  \includegraphics[width=0.8\linewidth]{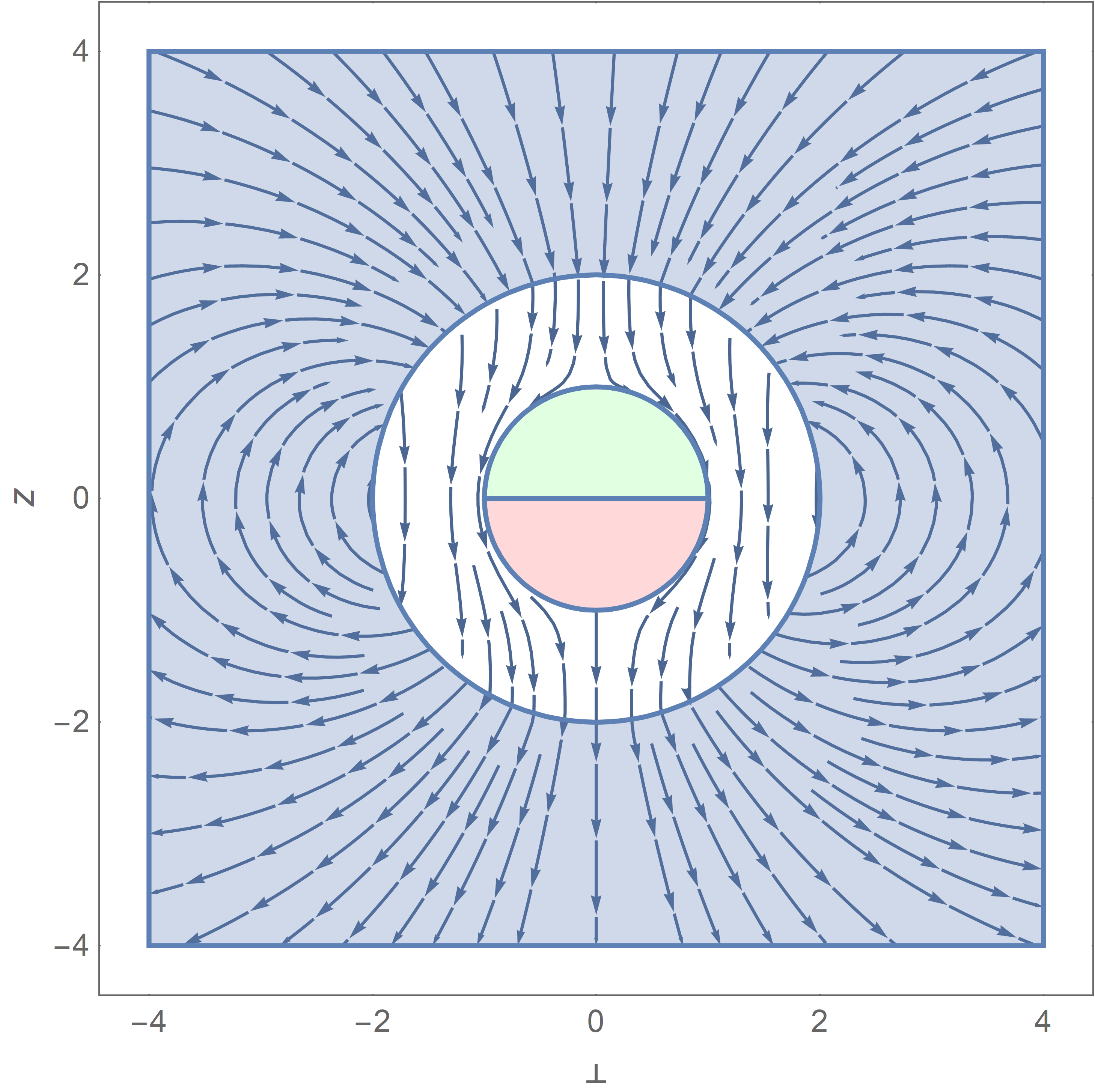}
  \caption{}
\end{subfigure}
\caption{Streamlines for $r_2\rightarrow \infty$, with   $a=1 \, \mu$m, $r_1=2 \, \mu$m and $\varepsilon=4$. Panel (a): the electric field. Panel (b): the magnetic field.}
\label{r2=inf}
\end{figure}

Here we keep the vacuum region between $a$ and $r_1$ but extend the TI to infinity. Going back to the system of equations we observe that 
 (\ref{radialphi21}), (\ref{2a}), (\ref{3a}) and (\ref{4a}) reduce to  
\begin{equation}
\varepsilon (l A_l^2 r_2^{l-1})=0,\qquad
A_l^2 r_2^l = 0,\qquad 
C_l^2 r_2^l= 0,\qquad 
l C_l^2 r_2^{l-1}=0,
\end{equation}
 since $r_2^{-n}=0$ for all $n \geq 1$ in the limit $r_2\rightarrow \infty$. So the only way that the above conditions  are satisfied is to take
$
A_l^2=C_l^2=0.
$
On the other hand, since region 3 is now equivalent to $r \rightarrow \infty$   the coefficients $B_l^3$ and $D_l^3 $ can be set equal to zero such that the potentials vanish. By eliminating these four variables $A_l^2, C_l^2, B_l^3, D_l^3 $ in Eqs. (\ref{A51}-\ref{A60}), a system of six equations is obtained. The solution is given by
\begin{eqnarray}
A_l^1&=&(l+1)(\varepsilon -1) \lambda^l_d,\qquad
B_l^1=-r_1^{2l+1}(l+\varepsilon + l\varepsilon) \lambda^l_d,\qquad
B_l^2=-(2l+1)r_1^{2l+1}\lambda^l_d, \nonumber \\
C_l^1&=&-\tilde{ \alpha}(l+1)\lambda^l_d,\qquad
D_l^1=-\tilde{ \alpha} l a^{2l+1} \lambda^l_d,\qquad
D_l^2=-\tilde{ \alpha} l (a^{2l+1}-r_1^{2l+1}) \lambda^l_d, \nonumber \\
&& \hspace{3cm}\lambda^l_d=\frac{a^{1+1}V_l}{a^{2l+1}(1+l)(\varepsilon-1)-r_1^{2l+1}   (l+\varepsilon(l+1))}.
\label{CASED}
\end{eqnarray}
The streamlines for  this case  are shown in Fig. \ref{r2=inf}. Its relation to the case C will be discussed in the next subsection.
\subsection{The relation between the cases C and D }
\label{case5}
The particular setups for Case C and Case D are shown in Figs. (\ref{r1=a}) and (\ref{r2=inf}), respectively. Such plots  suggest that, in our approximation,  both cases could be  linked by the transformations
\begin{equation}
\theta_1 \leftrightarrow \theta_2, \quad  ({\tilde \alpha} \rightarrow -{\tilde \alpha}) ,\qquad
r_1 \leftrightarrow  r_2,\qquad 
\varepsilon_1  \leftrightarrow  \varepsilon_ 2, \quad \left(\varepsilon \rightarrow {1}/{\varepsilon}\right),
\label{TRANS}
\end{equation}
where $\varepsilon=\varepsilon_1/\varepsilon_2$. 
Let us verify this assertion starting from Case C, where we perform the above transformations (\ref{TRANS}) in Eqs. (\ref{COEFF_CASE_C}) denoting  the resulting coefficients with a bar superindex. Explicitly we obtain
\begin{eqnarray}
{\bar \lambda}^l_c &=& -\varepsilon \lambda^l_d \\
{\bar A}_l^{2}&=&(l+1) (1/\varepsilon -1) (-\varepsilon \lambda^l_d)= A_l^1,\\
{\bar B}_l^{2}&=& r_1^{2 l+1} (l /\varepsilon +l+1)(-\varepsilon \lambda^l_d)= B_l^1,\\
{\bar B}_l^{3}&=&(2 l+1) (1/\varepsilon) r_1^{2 l+1}(-\varepsilon \lambda^l_d),=B_l^2\\
{\bar C}_l^{2}&=&\tilde{\alpha}  (l+1) (1/\varepsilon) (-\varepsilon \lambda^l_d)= C_l^1,\\
{\bar D}_l^{2}&=& \tilde{\alpha}  (l/\varepsilon)  a^{2 l+1}(-\varepsilon \lambda^l_d)=D_l^1,\\
{\bar D}_l^{3}&=&\tilde{\alpha}  (l/\varepsilon)  \left(a^{2 l+1}-r_1^{2 l+1}\right)(-\varepsilon \lambda^l_d)=D_l^2,
\end{eqnarray}
where the coefficients in the right hand side of the above equations are those of the Case D, given in Eq. (\ref{CASED}). 
However, the following  aspect   in the comparison  still requires clarification: in the $r_1=a$ configuration there are two MEP gradients, while in the  $r_2 \rightarrow \infty $ configuration there is only one. In the first case, one of them appears at the interface $r=r_2$, while the second corresponds to the surface of the conductor $r=a$, since $\tilde{ \alpha}=0 $ is assumed in its interior. However, the interface at the perfect conductor does not play any r\^ole  as previously stated.
A further remark is in order. In an abuse of notation we have denoted the relative permittivity in the cases C ($\varepsilon_C$) and D ($\varepsilon_D
$) by the same symbol $\varepsilon$. Since  $\varepsilon=
\varepsilon_2/\varepsilon_1$, this means that in fact we  have $\varepsilon_D= 1/\varepsilon_C$. Analogously, ${\tilde \alpha}_D= -{\tilde \alpha}_C$.
Summarizing, we have shown that the calculation of the coefficients for the case C in Eq.(\ref{COEFF_CASE_C}), after setting  $r_1=r_2$, in terms of the parameters corresponding to the case D, i. e. taking  $\varepsilon_C= 1/\varepsilon_D$  and   ${\tilde \alpha}_D= -{\tilde \alpha}_C$  yields the correct expressions for the coefficients of the case D in Eq.(\ref{CASED}). In other words,  when $r_1=r_2$ in the cases C and D related  according to  the Eq. (\ref{TRANS}), the final conclusion is that, at a given point,  the electric fields are equal in direction and magnitude in each of the setups, while the magnetic fields have the same magnitude but opposite directions.
\subsection{$r_1=a$ and $r_2\rightarrow \infty$}
\label{case6}
The solution is obtained taking $r_1=a$ in   Eq.(\ref{CASED}) and  it is 
\begin{eqnarray}
A_l^1&=& -\frac{a^{-l}(1+l)(\varepsilon-1) V_l}{2l+1},\quad
B_l^1= \frac{a^{l+1}(l+\varepsilon + l\varepsilon)V_l}{2l+1}, \qquad A_l^2=0, \qquad B_l^2= a^{l+1}V_l, \label{23} \nonumber \\ 
C_l^1&=&\frac{ {\tilde \alpha} a^{-l}(l+1)V_l}{2l+1},\qquad   D_l^1=\frac{{\tilde \alpha} a^{l+1} l V_l}{2l+1},  \qquad C_l^2=0, \qquad D_l^2=0. \label{24}
\end{eqnarray}
Note that  in the region $a < r < \infty$  the magnetic potential is null. This is what is expected for a homogeneous medium with constant $\tilde{ \alpha}$, after recalling that the discontinuity of the MEP in the perfectly conducting interface at $r_1=a$ does not contribute to the current producing the magnetic field .  On the other hand, although the coefficients $A_l^1$ and $B_l^1$ are not zero, the potential in the region 1  only makes sense  at $r=a$. It is expected, however, that $\Phi_1(a,\theta)=\Phi_2(a,\theta)$ satisfying the correct boundary conditions (\ref{BCPOT}). In fact, substituting the Eqs. (\ref{23}) we verify that 
\begin{eqnarray}
\Phi_1(a,\theta)&=&\sum_{l}{}^{\prime} \left[ \left( -\frac{a^{-l}(l+1)
(\varepsilon-1) V_l}{2l+1} \right) a^l + \left(\frac{a^{l+1}(l+\varepsilon + l\varepsilon)V_l}{1+2l}\right) a^{-(l+1)} \right] P_l(\cos \theta)\nonumber \\
&=&\sum_{l}{}^{\prime}\,  V_l P_l(\cos \theta).
\end{eqnarray}
This is an anticipated  result  because the above boundary condition was imposed from the very beginning to determine  $A_l^1$ and $B_l^1$.  
At the same time,
\begin{eqnarray}
\Phi_2(a,\theta)&=&\sum_{l}{}^{\prime} \left( B_l^2 a^{-(l+1)}\right) P_l(\cos \theta)
=\sum_{l}{}^{\prime}\, V_lP_l(\cos \theta).
\end{eqnarray}
The electric potential in the bulk ($a < r < \infty$) is given by 
\begin{eqnarray}
\Phi_2(r,\theta)&=&\sum_{l}{}^{\prime} \left( B_l^2 r^{-(l+1)} \right) P_l(\cos \theta)=\sum_{l}{}^{\prime} \left(\frac{a}{r}\right)^{l+1} V_l P_l(\cos \theta),
\end{eqnarray}
that once again coincides with the trivial case. As observed  at the end of section  \ref{case1} this result is independent of the permittivity which now is $\varepsilon=4$.

As an additional check of our results we verify that the non zero values of $C_l^1$ and $D_l^1$ in Eq.(\ref{24}) reproduce the boundary condition on the perfect conducting plates imposing that the normal component of the magnetic field is zero. This is given by Eq.(\ref{DRPSI}) evaluated at $r=a$, which  after substitution yields
\begin{equation}
\left(\frac{\partial \Psi_1}{\partial r}\right)_{r=a}=\sum_{l}{}^{\prime} \Bigg[ \left(\frac{ {\tilde \alpha} a^{-l}(l+1)V_l}{2l+1}\right) l  a^{l-1} - \left( \frac{{\tilde \alpha} a^{l+1} l V_l}{2l+1}\right) (l+1) a^{-(l+2)}\Bigg] P_l(\cos \theta)=0.
\end{equation}
As it was done in the  previous limiting cases, the coefficients of region 1 can be safely ignored, because those in the region 2 satisfy the boundary conditions at the plates of the semiespherical capacitor. Thus, it is clear that the present case is equivalent to the trivial one. 

In this section we have highlighted three particular cases, each representing a limiting case of the general setup. However, two of them ($r_1=a $ and $r_2=\infty $) are closely related , while the case F is experimentally difficult to achieve. Despite the usefulness of the graphic representations presented in this section, it should be noticed  that they do not intend to make explicit the magnitude of the fields produced, since they only show streamlines. 
With this in mind, in the next section we calculate the magnitudes of the fields and explore how to find the configurations that maximize them, bringing the problem closer to a possible experimental consideration.

The determination  of the field streamlines  is made by taking the potentials up to $l=9$. As will be seen in Section \ref{experimental} this approximation is more than enough for the configurations considered here.  The configuration with   $a=r_1$ will turn out to be  relevant  from the phenomenological point of view and  Fig. \ref{r1=a} shows this case, in which $r_1=\frac{1}{3}r_2$, a value that is within the domain that does not need large $l$  approximations.

\section{Numerical calculations of the magnetic field}
\label{experimental}

Having in mind a lowest attainable value for the magnetic field of the order of  $10^{-2}\,$G  we  fix the parameters that determine the magnetic field  (MEP, permittivity, radii), and estimate its magnitude at different points in space. 
Let us recall that we have taken  the  regions 1 and 3 to be the vacuum, so that the magnetoelectric effect arises only from the TI in the region 2.  It is important to focus our attention close to the external interface of the TI, because  this region  is  accessible to  measurement devices  and also takes advantage of  its proximity with the plates of the semiespherical capacitor which source the magnetic  field. As for the direction, we analyze the cases $\theta=0,\pi/4,\pi/2$. 

In particular  we will pay special attention to the case where the spherical shell touches de capacitor plates (i.e. $r_1=a$) and when the magnetic field is measured at the external interface (i.e. at $r=r_2$). Under these conditions the corresponding fields can be written as 
\begin{eqnarray}
\left[ r_{2}B_{r}\right] _{r_{1}=a,\;r=r_{2}} &=&\tilde{\alpha}%
\sum_{l}{}^{\prime} \,  F_{l}(s)l(l+1)V_{l}P_{l}(\cos \theta ),\nonumber \\\left[
r_{2}B_{\theta }\right] _{r_{1}=a,\;r=r_{2}}&=&-\tilde{\alpha}%
\sum_{l}{}^{\prime} \, F_{l}(s)lV_{l}\frac{dP_{l}(\cos \theta )}{d\theta }
\end{eqnarray}                        
with 
\begin{equation}
F_{l}(s)=\varepsilon \frac{s^{l+1}\left( 1-s^{2l+1}\right) }{\left[
(\varepsilon -1)(l+1)s^{2l+1}+(l\varepsilon +l+1)\right] }, \quad 
s =\frac{r_{1}}{r_{2}}.
\end{equation}
Dimensional reasons indicate  that the potentials $\left[ \Phi\right] _{r_{1}=a,\;r=r_{2}}$ and $\left[ \Psi \right] _{r_{1}=a,\;r=r_{2}}$, which are linear in $V_l$, are only functions of $s$.

For the following numerical  estimations, we set $\tilde{ \alpha}=\alpha\approx1/137$, the minimum value for a TI which we take  as ${\rm TlBiSe}_2$, with  $\varepsilon_2 \approx4$. Also we set  $V=3$ V;  $a,r_1$ of the order of  $\mu$m=$10^{-4}$ cm and we fix $r_2=1\, \mu$m. This choice of the TI together with the  characteristics of the setup are motivated by  Ref. \cite{esfera}.  We work in  
the CGS system where the electric field and the magnetic field are measured in ${\rm statV}/{\rm m}
$ and G, respectively. 
\begin{figure}[htb!]
\centering
\begin{subfigure}{.5\linewidth}
  \centering
 \includegraphics[width=1\linewidth]{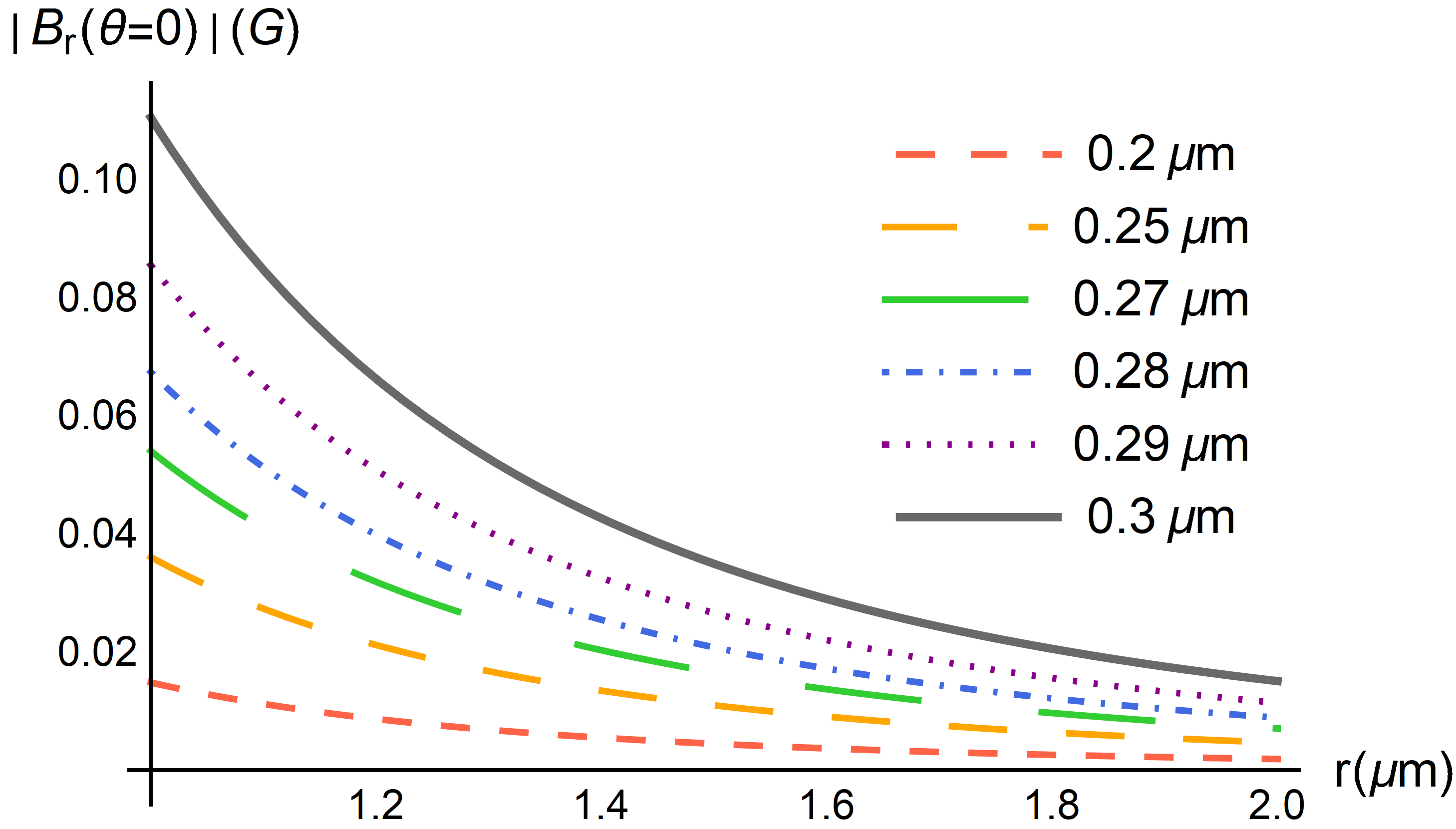}
  \caption{}
  \label{radial3}
\end{subfigure}%
\begin{subfigure}{.5\linewidth}
  \centering
  \includegraphics[width=1\linewidth]{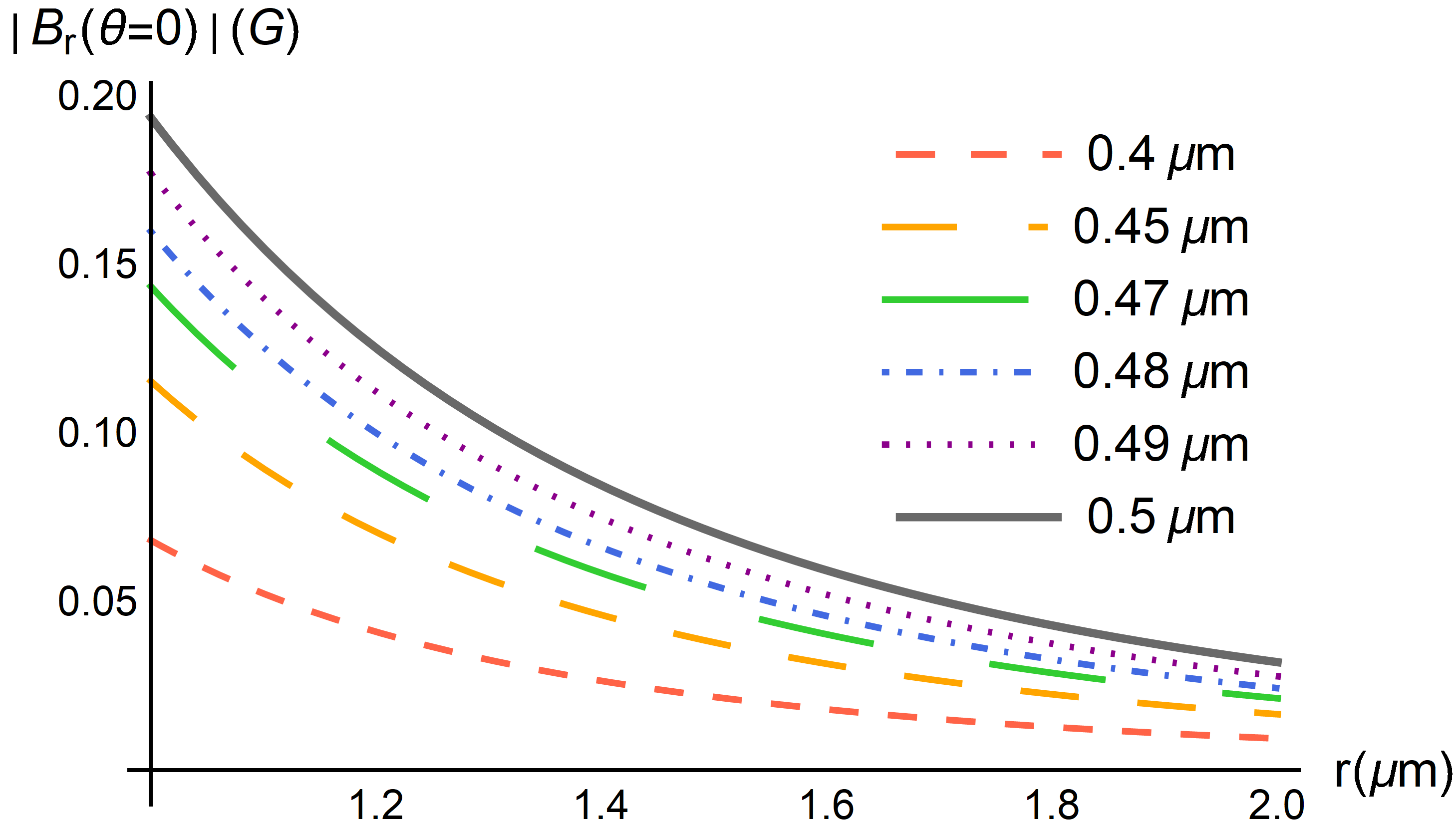}
  \caption{}
  \label{radial5}
\end{subfigure}
\begin{subfigure}{.5\linewidth}
  \centering
  \includegraphics[width=1\linewidth]{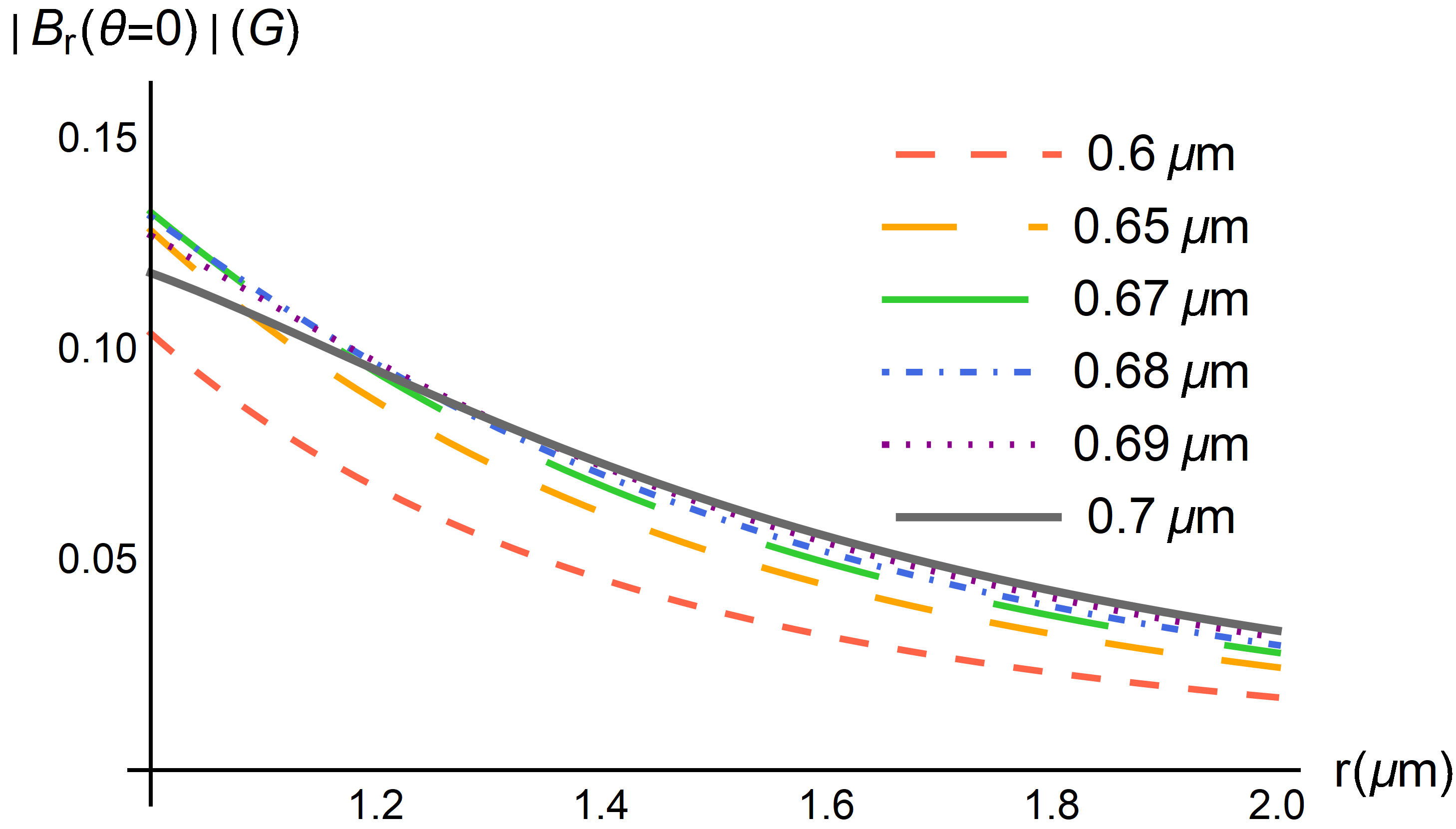}
  \caption{}
  \label{radial7}
\end{subfigure}%
\begin{subfigure}{.5\linewidth}
  \centering
  \includegraphics[width=1\linewidth]{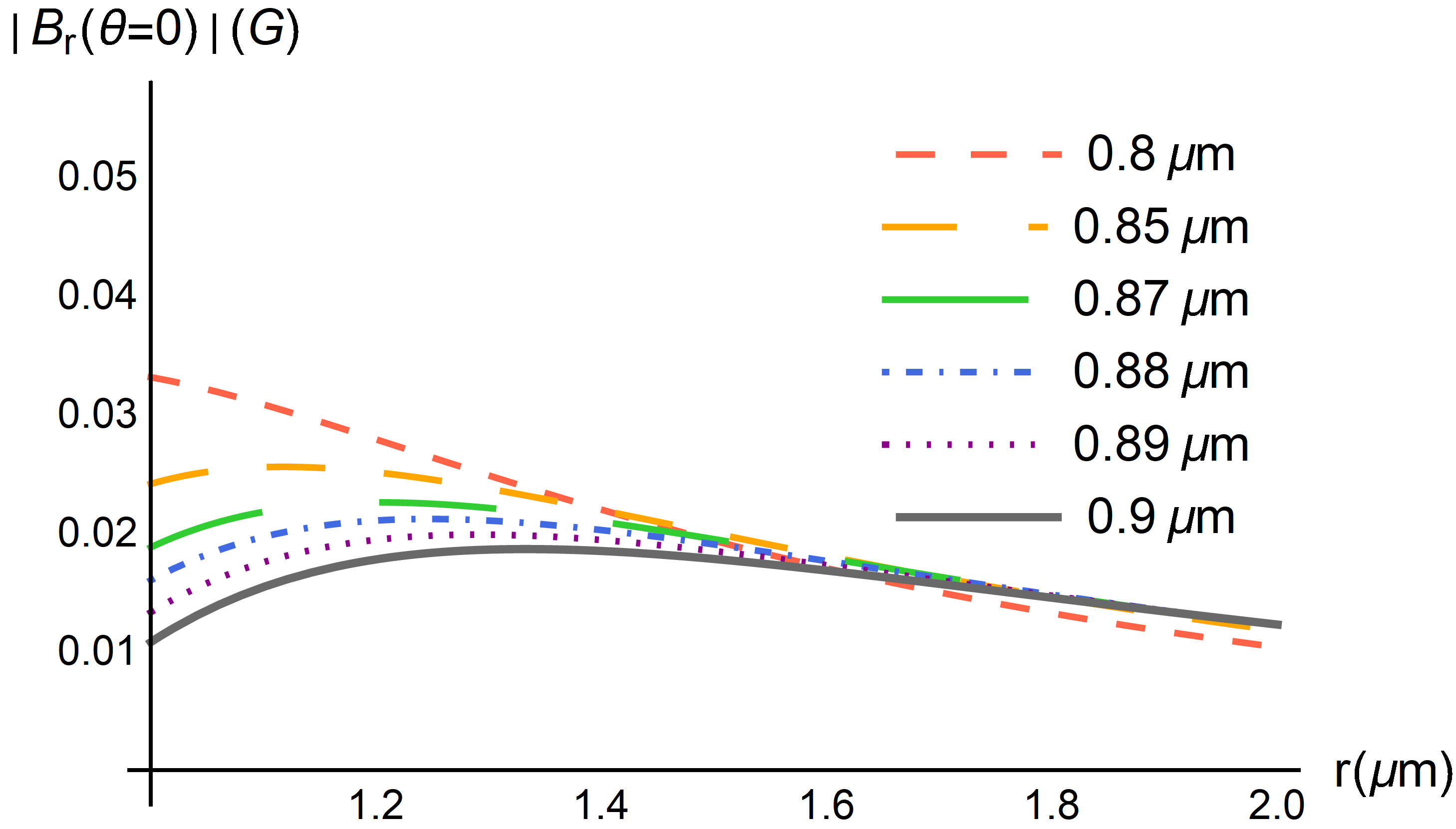}
  \caption{}
  \label{radial9}
\end{subfigure}
\caption{Plot of $|\mbf{B}(r_2, 0)|$ at $\theta=0$ and $r_2=1\, \mu$m, as a function of distance,  for different values of $a$. The values for $r_1$ are: Panel (a) $r_1=0.3 \, \mu$m, Panel (b) $r_1=0.5 \, \mu$m, Panel  (c) $r_1=0.7 \, \mu$m and Panel  (d) $r_1=0.9\, \mu$m.
In the panels (a) and (b), the choice $a=r_1$ maximizes the magnitude of the field near $r=r_2$. For values greater that $r_1\approx 0.6 \, \mu$m this is no longer the case, as shown in the panels (c) and (d).}
\label{radial}
\end{figure}
\subsection{Optimal configuration for $\theta=0$}
Examining the behavior of the magnetic field in the direction  $\theta=0$, where $\mbf{B}=B_r  \mbf{{ \hat r}}$, provides a general notion of the problem. After this, we make some estimations of the total magnitude of the field in other directions. As a first approximation, we assume that the magnetic field is maximized near the external interface for any configuration. This hypothesis is reasonable because the field must decrease with distance. Moreover, it can be shown that if $r_1 \leq 0.6 \, \mu$m, then $r_1=a$ maximizes $|\mbf{B}(r_2, 0)|$ in the external interface at $\theta=0$. Some examples are shown in Fig. \ref{radial} where $r_2=1\mu$m. Each panel of the figure corresponds to a given value of $r_1 < r_2$, with each set of curves  labeled by $a$ , where $a < r_1 < r_2$.   Notice that $a=r_1$ in fact provides a maximum value for   $|\mbf{B}(r_2,0)||$ for small values of $r_1$ (Fig. \ref{radial3} and Fig. \ref{radial5}). In contrast, for larger values of $r_1$ there is an optimal choice of $a\neq r_1$ for each particular case (Fig. \ref{radial7} and Fig. \ref{radial9}). To find this optimal value, one must solve the equation ${d|\mbf{B}(r_2, 0)|}/{da}=0$ at $r=r_2$, for fixed $r_1$ and $r_2$.

\begin{figure}[htb!]
\includegraphics[width=0.5\linewidth]{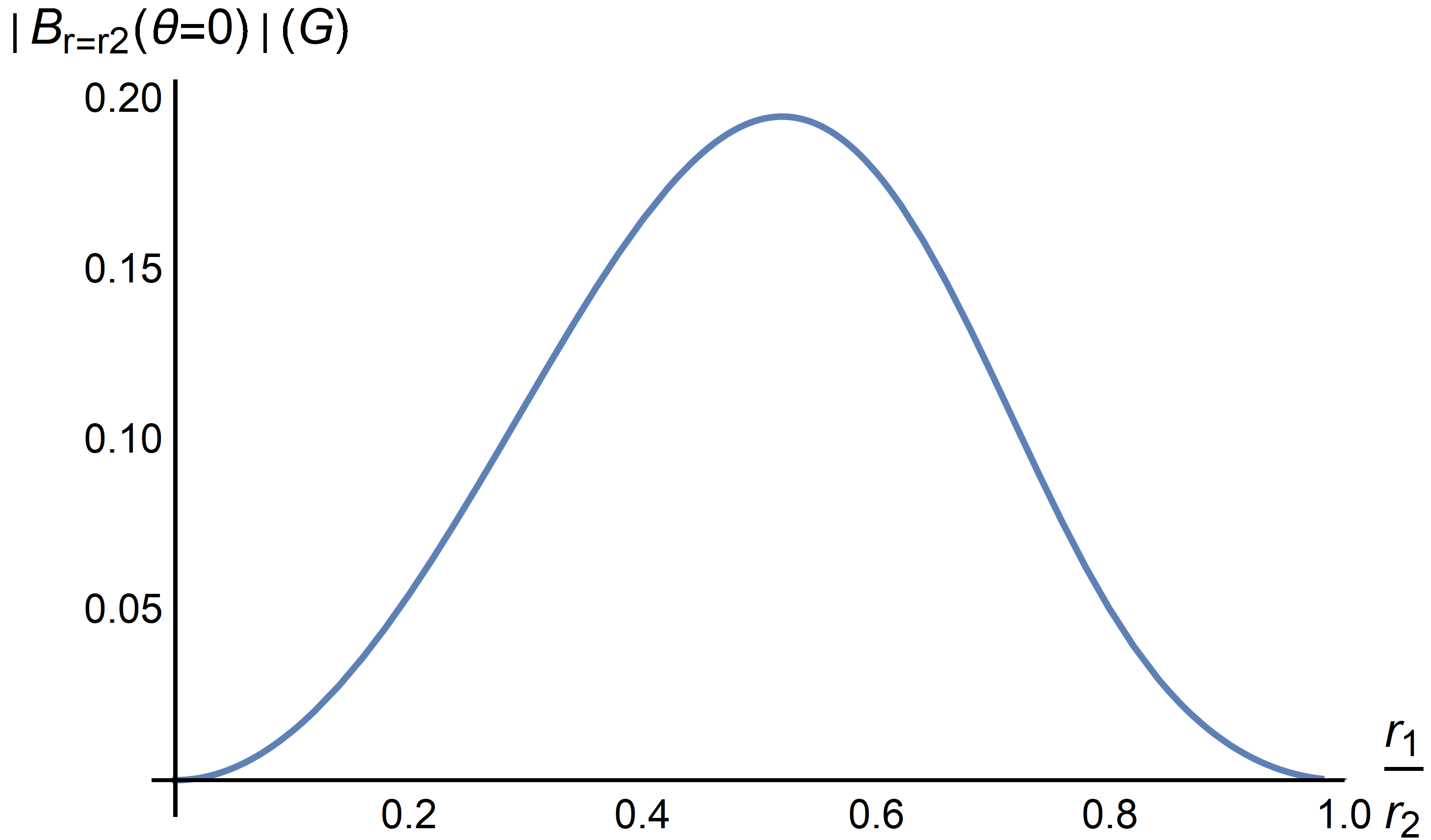}
\caption{Plot of $|\mbf{B}(r_2, 0)|$ at $\theta=0$, as a function of $r_1$, ($ 0< r_1 < 0.95 \,\mu$m), at the interface $r=r_2=1 \mu$m, when $a=r_1$. The magnetic field reaches its maximum value ($0.2G$) at $r_{1 \rm{m}} \approx 0.5 \, \mu$m.}
\label{maximize}
\end{figure}

Continuing with this analysis, it should be noted that given $r_2=1 \,\mu$m and choosing $a=r_1$, it is possible to determine the value of $r_{1{\rm m}}$ that maximizes $|\mbf{B}(r_2, 0)|$ at the external interface. One must simply solve the equation ${d|\mbf{B}(r_2, 0)|}/{dr_1}=0$. Fig. \ref{maximize} shows a plot of $|\mbf{B}(r_2, 0)|$ as a function of $r_1$, that reaches its maximum at $r_{1{\rm m}} \approx 0.5\, \mu$m. 
Regarding these observations, some comments can be made. Firstly, despite that in the domain $r_1 \geq 0.6 \, \mu$m the hypothesis that $a=r_1$ maximizes $|\mbf{B}|$ at the external interface is not satisfied, in general (for any value of $a$) the maximum of $|\mbf{B}|$ in that domain is much smaller than $0.2 \,$G, which is reached in the case where   $r_{1{\rm m}}\approx 0.5 \, \mu$m. Secondly, it is easier to make a coating of the magnetoelectric material on the surface of the conducting sphere than to leave an empty space between the capacitor plates  and the TI shell, so configurations for which $a$ equals $r_1$ will be  the most relevant. Since we do not want the magnetoelectric effect to disappear, in this and subsequent cases we  keep  the TI shell with a minimum thickness which we choose as  $0.05 \, \mu$m. In other words we  will explore $r_1$ in the range  $0<a=r_1 < 0.95\, \mu$, as shown in Fig.\ref{maximize}. {Also this constraint will force  us to stay away   from the dangerous points $a=r_1 \rightarrow r_2$,  when observing at $r=r_2$ and $ \theta= \pi/2 $.} From the previous discussion, it is concluded that the condition  $a=r_1 \approx 0.5 \, \mu$m gives rise to the most intense magnetic field in the direction of the z-axis. As seen in the  Figs. \ref{radial5} and \ref{maximize}, this field would be of order $0.2G$ at the interface $r=r_2$. 

\begin{figure}[htb!]
\centering
\begin{subfigure}{.5\linewidth}
  \centering
  \includegraphics[width=1\linewidth]{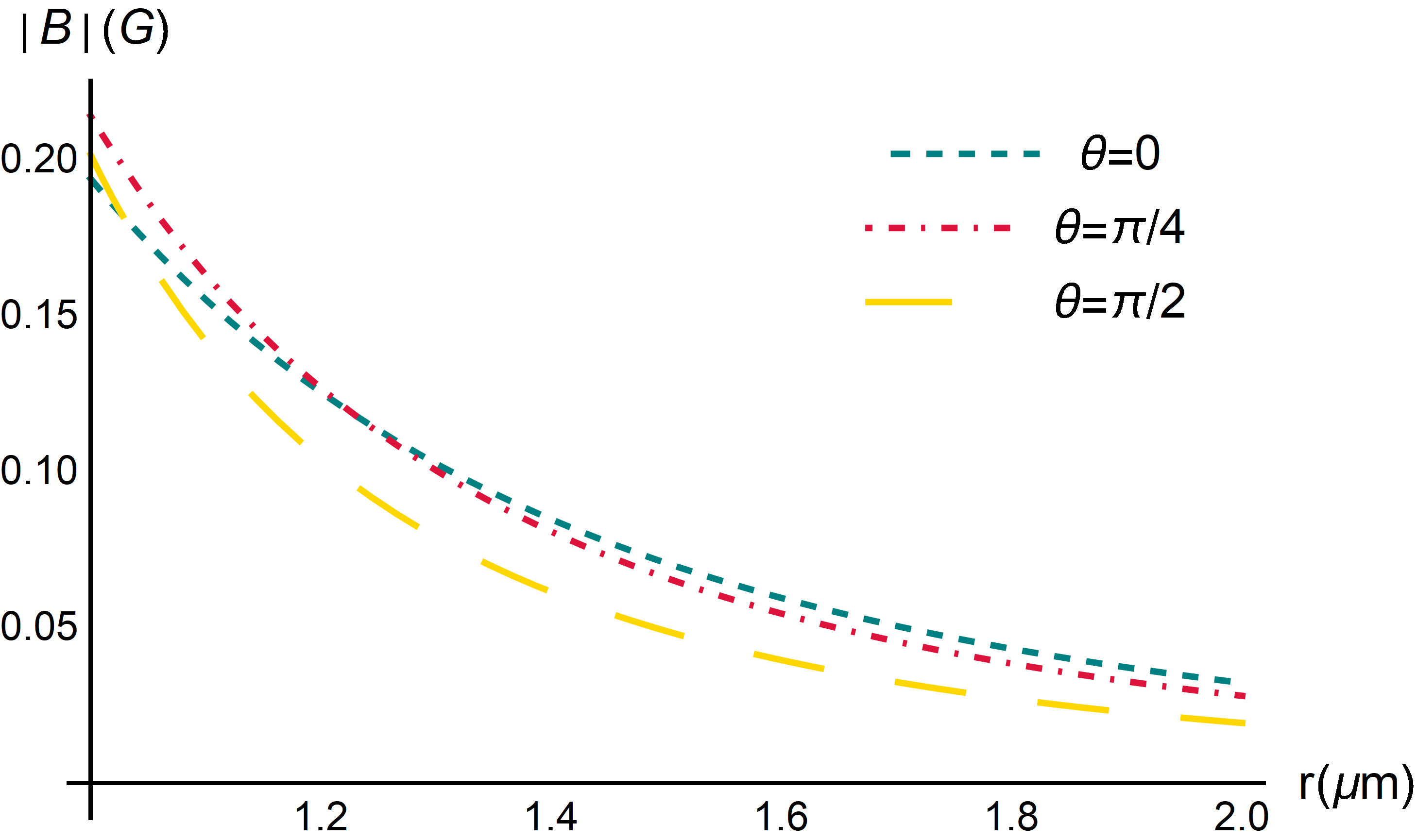}
  \caption{}
  \label{total5_}
\end{subfigure}%
\begin{subfigure}{.5\linewidth}
  \centering
  \includegraphics[width=1\linewidth]{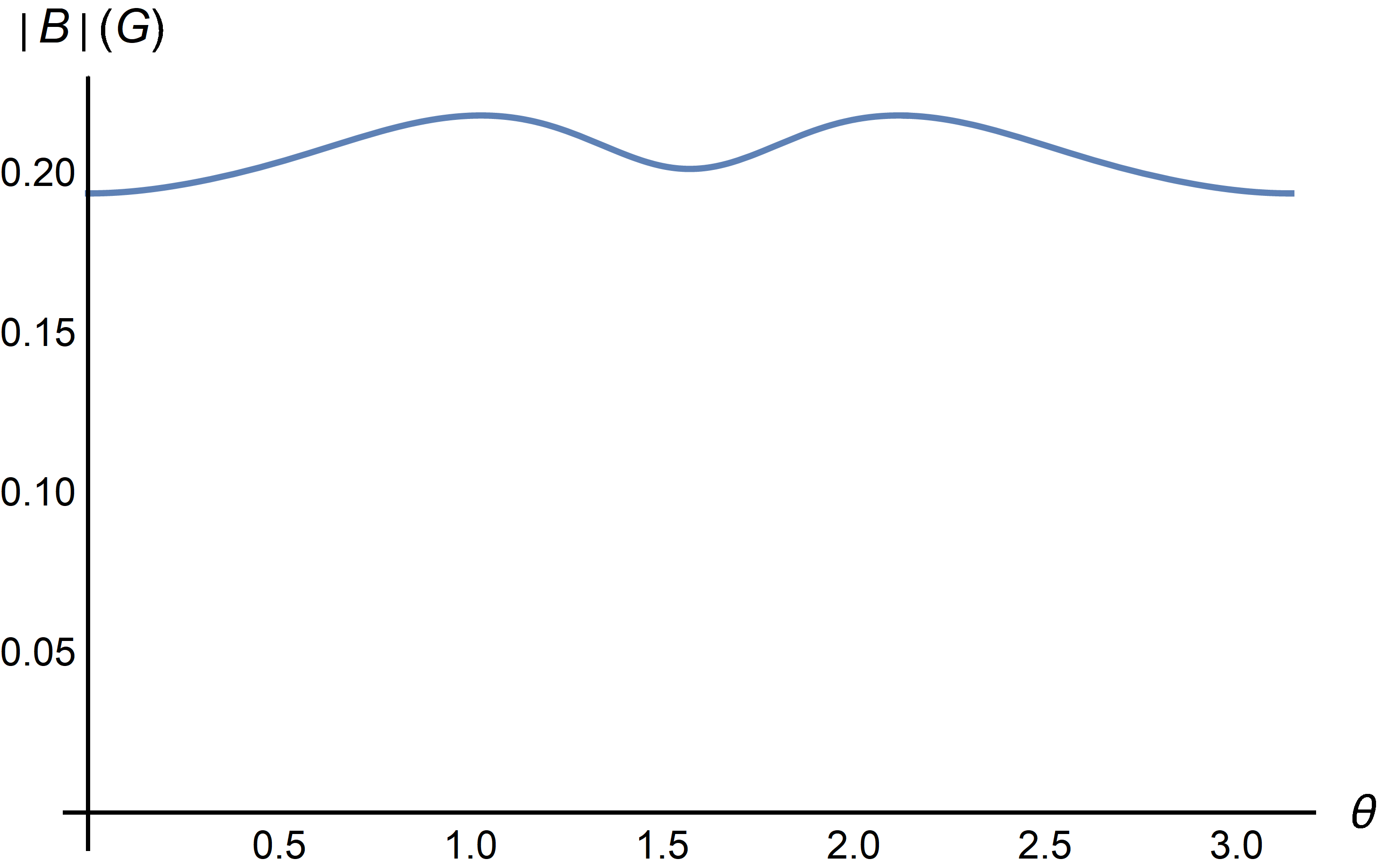}
  \caption{}
  \label{total5_theta}
\end{subfigure}
\caption{Panel (a): plot of the magnitude of the magnetic field in  the directions $\theta=0,\pi/4$ and $\pi/2$, as a function of distance. Panel (b): plot of  the magnitude of the magnetic  field on the interface $r=r_2$ as a function of $\theta$. The parameters are  $r_2=1 \,\mu {\rm m}$ and  $a= r_1=0.5 \, \mu{\rm m}$.}
\label{magnitudtotal}
\end{figure}

For completeness, it is important to know how the magnetic field behaves in other directions. Fig. \ref{total5_} shows a plot of the magnitude of the field $\mathbf{B}$ in the directions $\theta=0,\pi/4,\pi/2$ as a function of $r \geq r_2=1 \, \mu$m, choosing $a=r_1=0.5 \, \mu$m. Also, Fig. \ref{total5_theta} shows the total magnitude of the field in the interface $r=r_2$ as a function of $\theta$ for the configuration $a=r_1=0.5 \,\mu$m. It should be noted that, in this case, the magnitude of the field is highly isotropic.

\begin{figure}[htb!]
\includegraphics[width=0.5\linewidth]{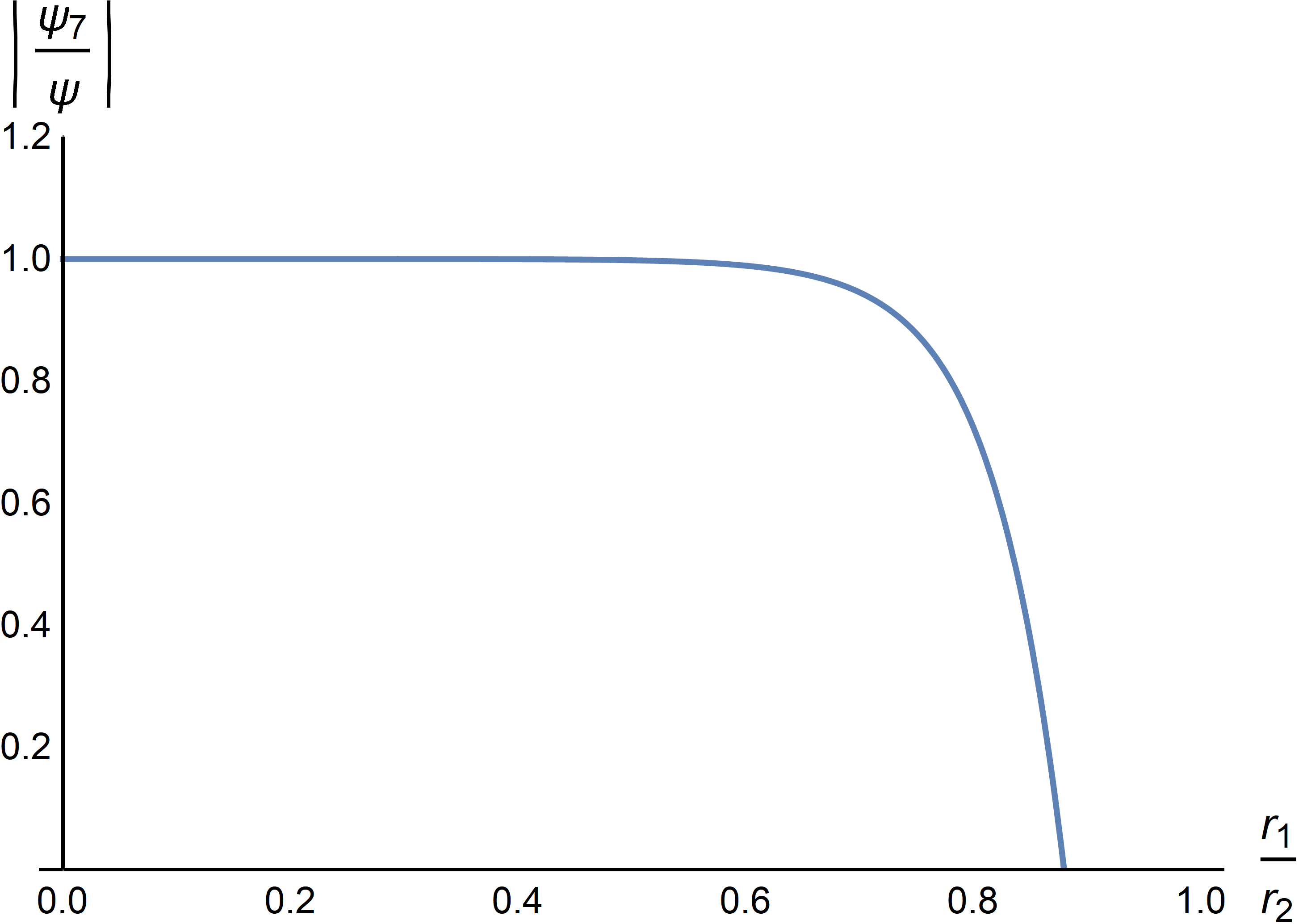}
\caption{Plot of $|\Psi_7/ \Psi|$, for $a=r_1$, $\theta=0$ and $r=r_2$, as a function of $r_1/r_2$. The approximation to order $l=7$ is valid up to $r_1 \approx 0.6 r_2$. To describe the behavior of the system for larger values of $r_1/r_2$ it is necessary to increase the order of the approximation.}
\label{approx}
\end{figure}

Finally, we comment on the precision required  in the calculations to adequately describe the physics associated with $r_2= 1 \, \mu$m and  large values of $r_1$ (as $r_1=0.9 \, \mu$m for example), where it is necessary to consider a great number of terms for the magnetic potential. The  Fig. \ref{approx} shows the quotient between the approximation of $\Psi$ at order $l=7$ and at order $l=1000$ for $r_1=a$ at  $\theta=0$ as a function of $r_1/r_2$.  For $r_1 \approx 0.7 \, r_2$, $l=7$ is no longer a good cut-off value to describe the system. For greater values of $r_1/r_2$, the approximation requires higher values of $l$.  It is important to clarify that to plot  Fig. \ref{radial3} and Fig. \ref{radial5}  it was enough to take $l=7$, while for Fig. \ref{radial7} and Fig. \ref{radial9} it was necessary to increase the number of terms.
\begin{figure}[htb!]
\includegraphics[width=0.5\linewidth]{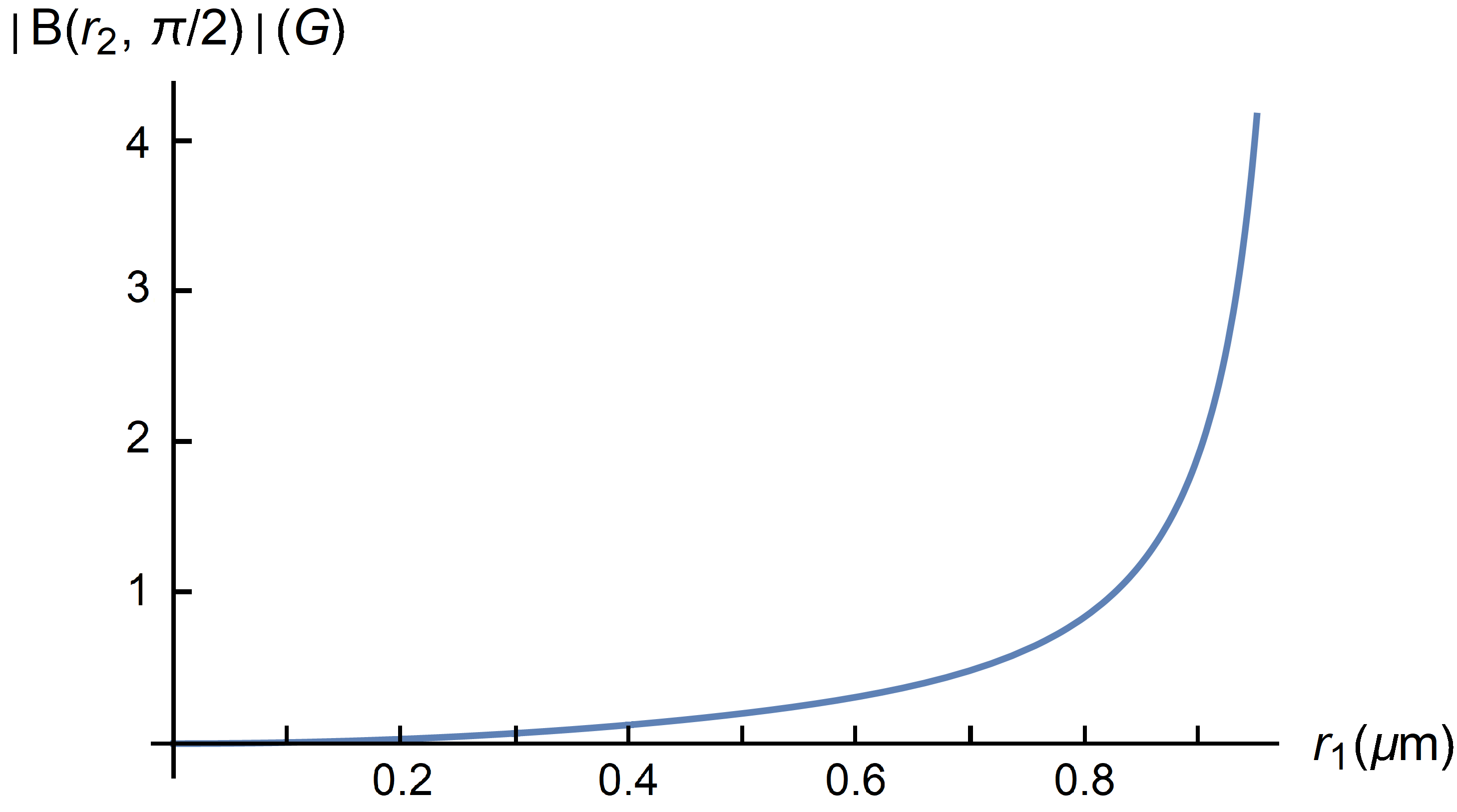}
\caption{Plot of $ |\mbf{B}(r_2, \pi/2)|$  at $r_2= 1 \, \mu$m, as a function of $a=r_1$, for $\theta=\pi/2$.}
\label{maximizepi2}
\end{figure}
\subsection{Optimal configuration for $\theta=\pi/2$}
\begin{figure}[htb!]
\centering
\begin{subfigure}{.5\linewidth}
  \centering
\includegraphics[width=1\linewidth]{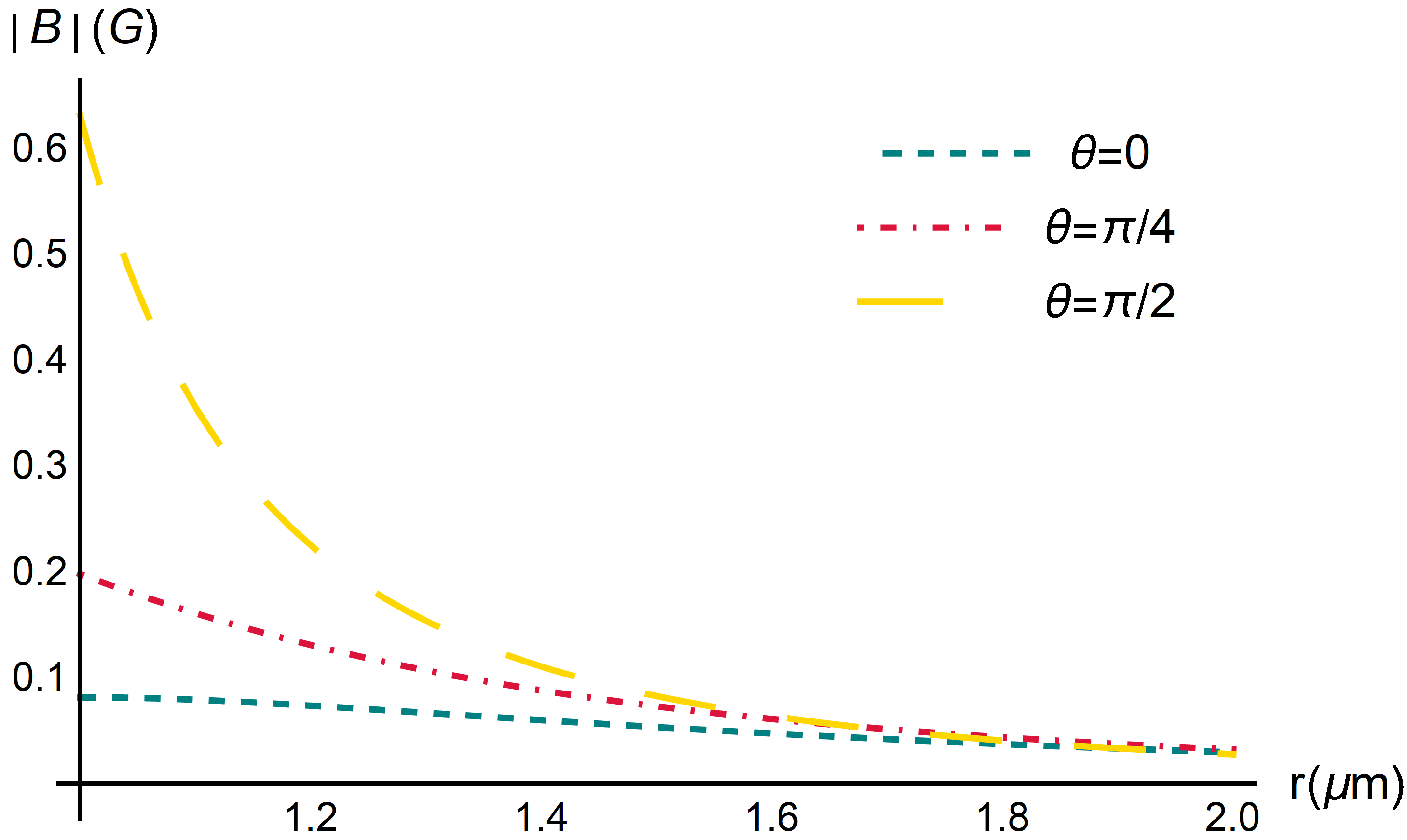}
  \caption{}
  \label{total75_}
\end{subfigure}%
\begin{subfigure}{.5\linewidth}
  \centering
 \includegraphics[width=1\linewidth]{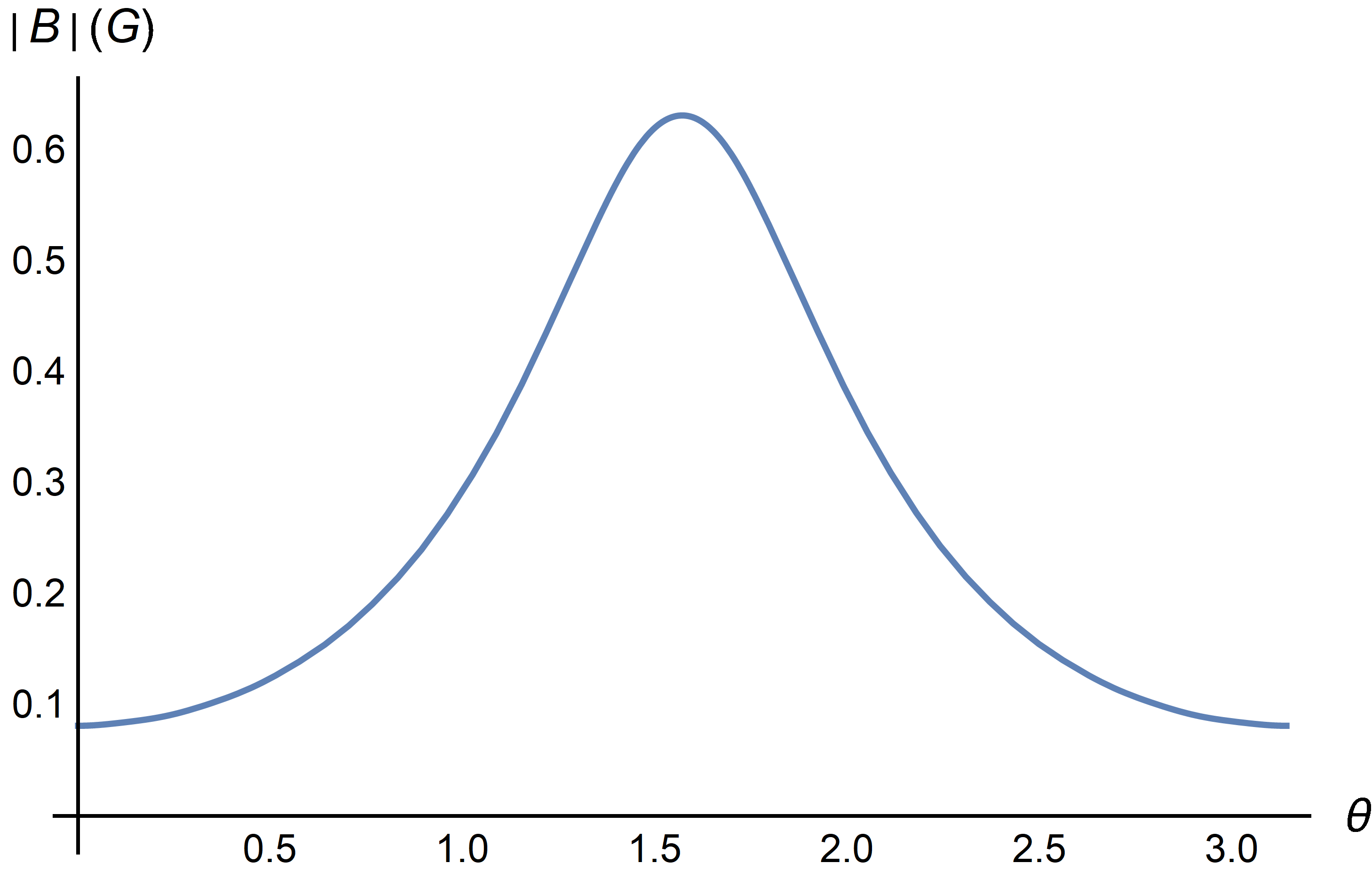}
  \caption{}
  \label{total75_theta}
\end{subfigure}
\caption{Panel (a): plot of the  magnitude of the magnetic field in the  directions $\theta=0,\pi/4$ and $\pi/2$, as a function of distance. Panel (b): plot of the magnitude of the magnetic field on the interface $r=r_2$ as a function of $\theta$. The parameters are  
$r_2=1 \, \mu$m and  $a=r_1=0.75\, \mu$m.}
\label{total75}
\end{figure}

 At these points the  total magnetic field is in the  direction of $\gv{\hat \theta}$. Making an analysis similar to that of the previous section, in Fig. \ref{maximizepi2} we show that, unlike the case  in $\theta=0$ (Fig. \ref{maximize}), the magnetic field in $\theta=\pi/2 $  at $r=r_2$ increases with $r_1$ when $a=r_1$. In this way, it is possible to generate large magnetic fields in the vicinity of $ \theta=\pi/2 $ that decrease dramatically in other directions. When $a=r_1=0.95 \, \mu$m we have  $|\mbf{B}(r_2, \pi/2)|=4.17\, $G at the external interface $r_2=1 \, \mu$m. The Fig. \ref{total75_} shows a plot of the  magnitude of the field with $r_1=0.75 \, \mu$m for three angles ($\theta=0, \pi/4, \pi/2) $ as a function of distance, while Fig. \ref{total75_theta} shows a plot of the field at the interface $ r=r_2 $ as a function of $\theta$. 
A first conclusion that suggests two distinct and even opposite empirical approaches is that the $a=r_1=0.75 \, \mu$m configuration (Fig. {\ref{total75_theta}) would generate intense but anisotropic fields, while the configuration
$a=r_1=0.5 \, \mu$m (Fig. \ref{total5_theta}) would generate almost isotropic fields but of lesser magnitude. {In the case $\theta=\pi/2$ it is particularly important to stay away from the points $a=r_1 \rightarrow r_2$, which we safely  avoid with the constraint $a=r_1 < 0.95 \, \mu$m $ < r_2= 1 \, \mu$m. }
\begin{figure}[htb!]
\includegraphics[width=0.5\linewidth]{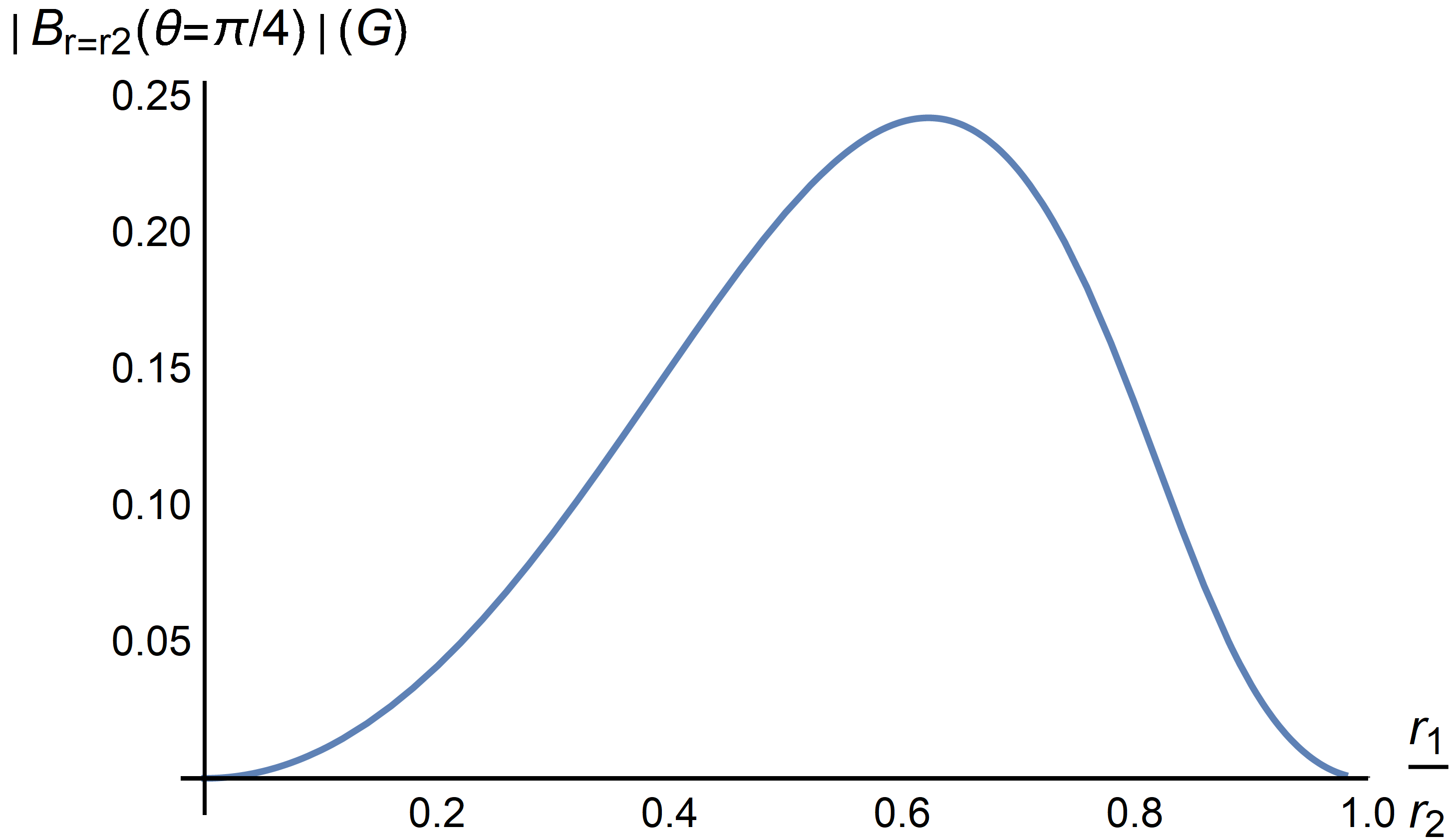}
\caption{Plot of $|\mbf{B}(r_2,\pi/4) |$  at $r_2= 1 \, \mu$m, as a function of $a=r_1$, for $\theta=\pi/4$.}
\label{maximizepi4}
\end{figure}
\subsection{Optimal configuration for $\theta=\pi/4$}
Similarly to the previous cases, it is possible to find the condition that maximizes the field in the direction $\theta=\pi/4 $. The Fig. \ref{maximizepi4} is a plot of $|\mathbf{B}|$ at $r_2= 1 \, \mu$m, as a function of $a=r_1$, for $\theta=\pi/4$  
 The maximum appears at $r_{1 \rm{m}}\approx 0.62 \,\mu$m. This case is a hybrid between the two previous ones, because it gives rise to fields nor as intense as in the case where  $r_{1 \rm{m}}=0.75 \, \mu$m, neither as isotropic as in the case where  $r_{1 \rm{m}}=0.5 \, \mu$m. The  Fig. \ref{total62_} shows the magnetic field as a function of $r$ produced by this configuration in the directions $\theta=0,\pi/4,\pi/2$. Likewise, the  Fig. \ref{total62_theta} shows the magnetic field at the external interface as a function of $\theta$.
 \begin{figure}[htb!]
\centering
\begin{subfigure}{.5\linewidth}
  \centering
\includegraphics[width=1\linewidth]{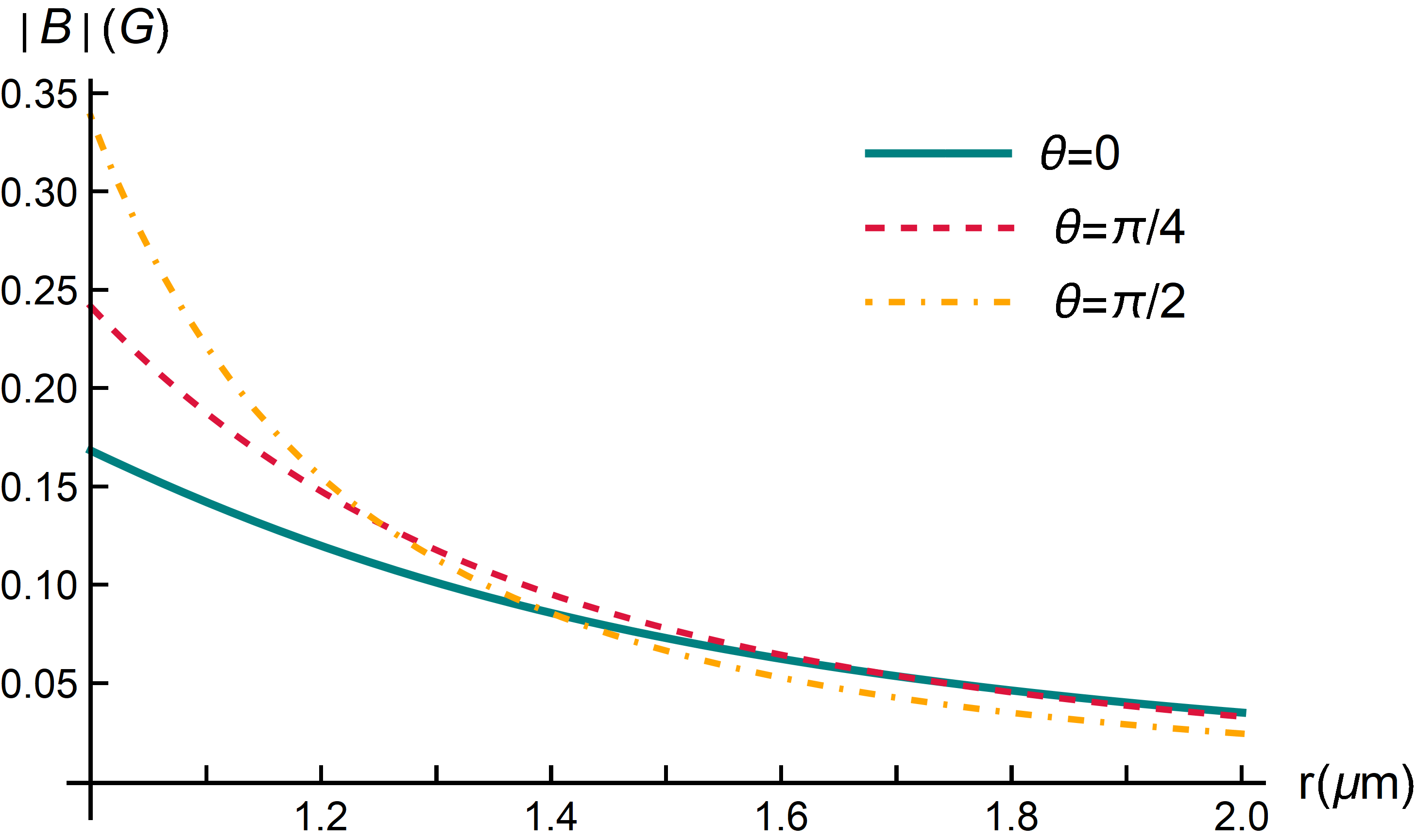}
  \caption{}
  \label{total62_}
\end{subfigure}%
\begin{subfigure}{.5\linewidth}
  \centering
 \includegraphics[width=1\linewidth]{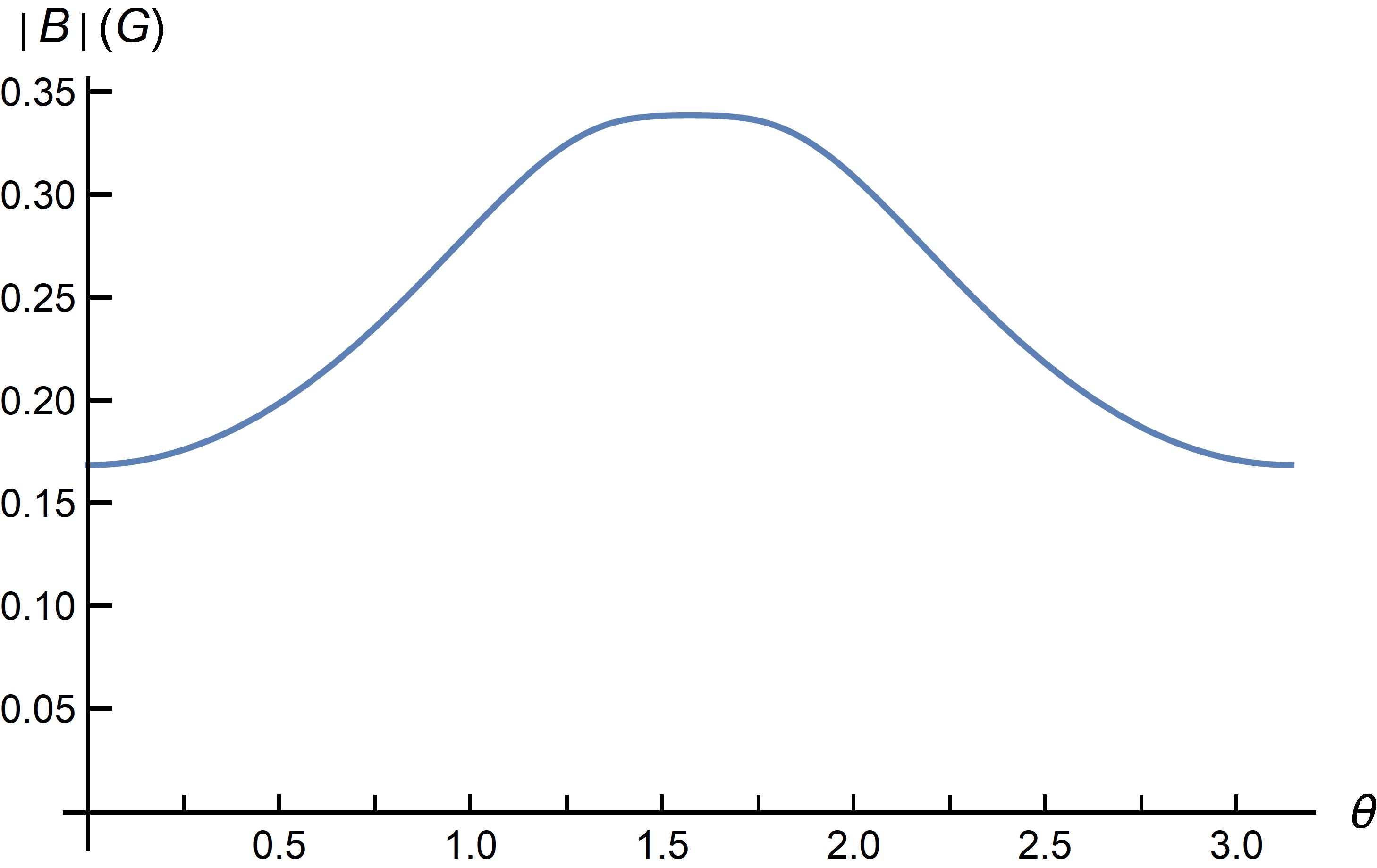}
  \caption{}
  \label{total62_theta}
\end{subfigure}
\caption{Panel (a): plot of the  magnitude of the magnetic field in the directions $\theta=0,\pi/4$ and $\pi/2$, as a function of distance. Panel (b): plot of the  the magnitude of the field on the interface $r=r_2$ as a function of $\theta$. The parameters are $r_2=1 \, \mu$m and  $a=r_1=0.62 \, \mu$m.}
\label{total62}
\end{figure}
We close this section with a general comment regarding the scaling of our setup as a whole, assuming   that $a, r_1, r_2$ are each  multiplied by a  factor $\Lambda >1$. Let us focus on the exterior region 3, where $ D_l^3 \rightarrow \Lambda^{l+1} D_l^3$ yielding 
\begin{eqnarray} 
\tilde{B}_{r}(r) &=&\sum_{l}{}^{\prime} \, \Lambda^{l+1}\left( \frac{D_{l}^{3}}{%
r^{l+2}}\right) (l+1)P_{l}(\cos \theta ), \\
\tilde{B}_{\theta }(r) &=&-\sum_{l}{}^{\prime} \, \Lambda ^{l+1}\left( \frac{%
D_{l}^{3}}{r^{l+2}}\right) \frac{dP_{l}(\cos \theta )}{d\theta }.
\end{eqnarray}
Here $\tilde B$ denotes the magnetic field after scaling the setup. If we now evaluate the fields produced by the scaled setup  at the scaled point ${\tilde r}= \Lambda r$ we obtain ${\tilde B}({\tilde r})=B(r)/\Lambda$. In particular this means that the magnetic field measured at the exterior interface of the scaled setup is reduced  by a factor $\Lambda $ with respect to that measured in the same interface before the scaling has been done. 
It is notable that an increase in system dimensions up to ten times, for example, starting  with an initial  optimum value of $|\mbf{B}| \approx 0.2$ G at the exterior interface, would maintain the possibility of experimentally measuring a detectable field of $\approx 0.02$ G at the new interface, for any of the two most significant configurations.

\section{Summary and discussion}

The electromagnetic response of TIs is described by a modification of Maxwell equations that includes  the production of  polarization (magnetization) due to the presence of magnetic (electric) fields. This phenomenon, called the magnetoelectric effect, occurs due to Hall currents at the interfaces, induced by gradients of  the  magnetoelectric polarizability $\vartheta$, an additional  topological  parameter  characterizing TIs,  besides their permittivity and permeability.  Although for several decades the magnetoelectric effect was known in normal magnetoelectric media, it was not until a few years ago that its study gained a remarkable strength, both theoretically and experimentally, due to its discovery in topological phases. Since then, significant advances have been made that may allow technological applications in subsequent years fostered by the production of composite materials with stronger magnetoelectric couplings.

In this work, we studied the magnetoelectric effect produced by  a capacitor with semispherical plates at  potentials $+V$ and $-V$, respectively, placed in vacuum ($\vartheta_1=0$) and  surrounded by a thick shell of a topological insulator with $\vartheta_2=\pi$, as shown in Fig. \ref{arreglo}. Using spherical coordinates appropriate for the symmetry of the problem, the static magnetic and electric fields were determined in the absence of additional free sources. The bulk regions are governed by the standard Maxwell equations for material media, while the effects of the topological insulator show up only at the interfaces  through the  boundary conditions (\ref{cb}) that result from the modified constitutive relations (\ref{consteq}). Our main concern is the magnetic field produced by the electric configuration. 
The general solution is difficult to handle, so we considered  a series expansion to first-order  in the parameter $\tilde{ \alpha}=(\vartheta_2\alpha)/\pi= \alpha$, where $\alpha $ is the fine structure constant. This expansion is justified only for media with a value of $\vartheta_2$ of order unit; however, it is a good starting point for further studies. We found that, for the most general configuration, the system results in the production of a magnetic field throughout the space.

\begin{table}[htb!]
\setlength{\tabcolsep}{10pt}
\renewcommand{\arraystretch}{1.5}
\begin{tabular}{|c|c|c|c|}
\hline
$a (\mu m)$ & $\left|\mathbf{B}(r=r_2,\theta=0)\right|(G)$ & $\left|\mathbf{B}(r=r_2,\theta=\pi/4)\right|(G)$ & $\left|\mathbf{B}(r=r_2,\theta=\pi/2)\right|(G)$ \\ \hline
$0.50$       & $\approx 0.20$            & $\approx 0.20$                & $\approx 0.20$                \\ \hline
$0.62$      & $\approx 0.18$          & $\approx 0.25$               & $\approx 0.35$               \\ \hline
$0.75$      & $\approx 0.10$            & $\approx 0.20$                & $\approx 0.60$                \\ \hline
\end{tabular}
\caption{Magnitudes of the magnetic field at the external interface ($r_2= 1 \, \mu$m) in the  directions $\theta= 0, \pi/4,\, \pi/2$ for the configurations $a=r_1= 0.50, \, 0.62, \, 0.75 \, \mu$m. }
\label{table_B}
\end{table}

Subsequently, the problem was particularized for some limiting cases, allowing a more direct interpretation of the results. On the other hand, this gave the opportunity to verify if the results obtained were correct, by reducing them to known cases. In addition, the streamlines of the electric and magnetic fields were plotted for different configurations.

Also, an important consideration was given to the experimental possibilities of the system by looking into various  configurations, keeping in mind that a good magnetometer, like a nitrogen-vacancy center
inside a diamond nanocrystal for example,  can detect magnetic fields in the range of $10^{-2}$ to $10^{+2}\, $G. Those configurations that produce the most intense fields were sought. In particular, by taking  $a=r_1$,  we emphasized the cases where the TI is in direct contact with the capacitor thus avoiding a vacuum region between them. This allows a stronger magnetoelectric effect as well as simplifies a possible construction  of the setup. Since the magnetic field decreases with distance we also concentrated in the results at the external interface located at  $r_2$. Fixing $r_2= 1\,\mu$m we are left with two variables which we explore: $ a=r_1$ and $\theta$. For a given direction, we plotted the magnetic field as a function of $a=r_1 < 0.95 \,\mu$m  finding a maximum which  increases in  value and occurs for larger values of $r_1$ as $\theta$ approaches  $\pi/2$.

{As repeatedly mentioned along the text, the discontinuity of the source potential at $r=a$ and $\theta=\pi/2$ makes these points unreliable in a numerical approximation. This is particularly  noticeable when considering $a=r_1$ and observing the magnetic field at $r_2$ in the limit of a thin TI shell, i. e. when $r_1 \rightarrow r_2$.} The calculation becomes particularly involved in this limit and requires more sophisticated numerical techniques which are out of the scope of the present work. However, according to the physics of the problem we expect that in the limit $a=r_1 \rightarrow r_2$, that is to say  when the TI disappears, the magnetic field should be zero everywhere, in particular at the external interface. This behavior, which is evident in the Figs. \ref{maximize} and \ref{maximizepi4},  is  consistent with  the cases when $\theta$ is $ \pi/4, \, 5\pi/16, \, 3\pi/8, \, 7\pi/16 $, approaching $\pi/2$, shown in the Fig. \ref{plots_dif_thetas}.   Nevertheless, this is not the case  in the Fig. \ref{maximizepi2} for $\theta\equiv \pi/2$. Thus,  our calculations are not trustworthy in this particular limiting case, which we avoid by restricting ourselves to $a=r_1 \leq 0.95 \, \mu$m in all the relevant configurations.
\begin{figure}[htb!]
\includegraphics[width=0.6\linewidth]{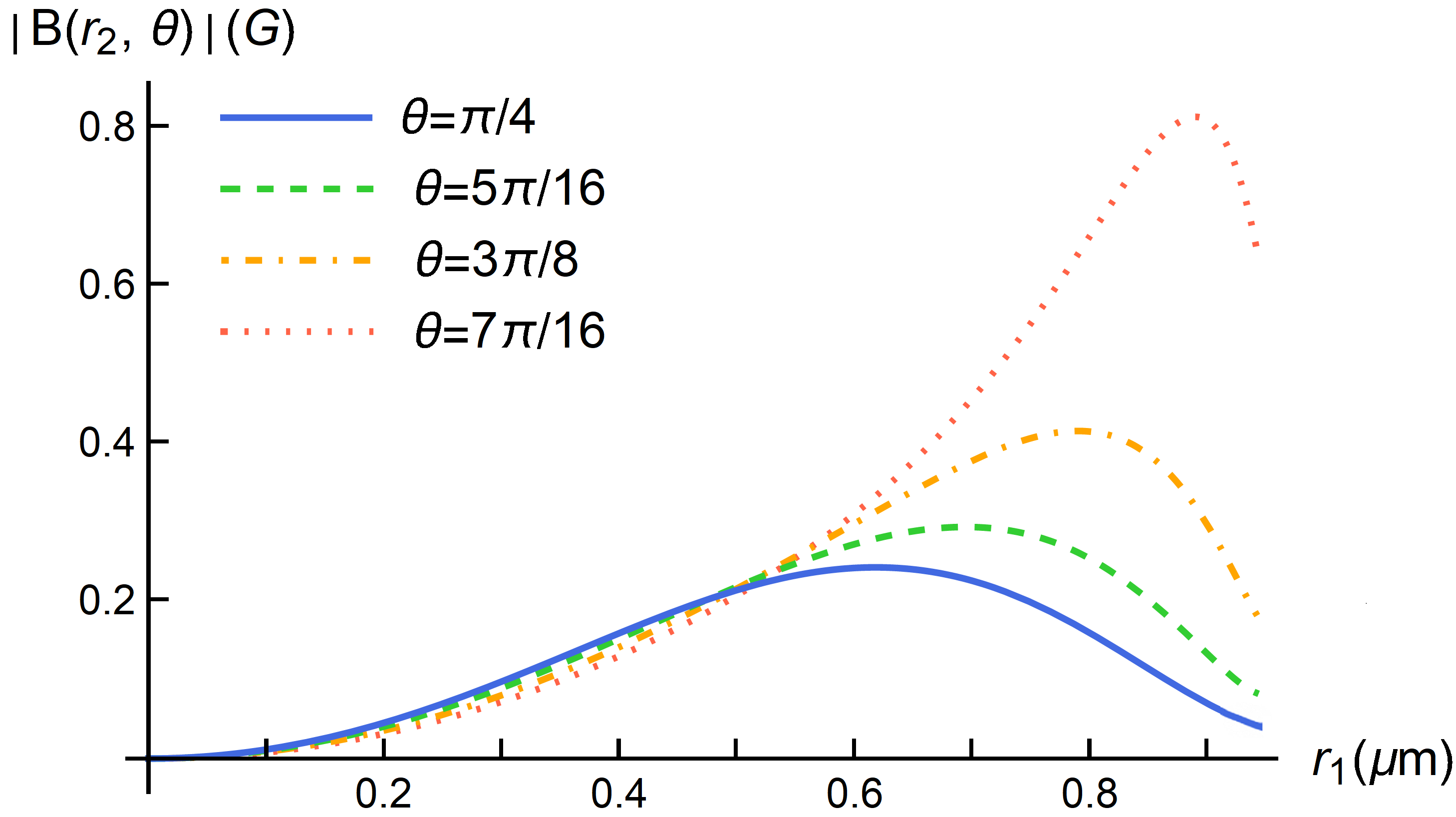}
\caption{Plot of $|\mbf{B}(r_2, \theta)|$ as a function of $a=r_1$, for different values of $\theta$ approaching $\pi/2$. The change in slope after the maximum is evident, which is compatible with  the correct zero  limit at $r_1=r_2=1 \,\mu$m.   }
\label{plots_dif_thetas}
\end{figure}
Within this limit we focus on the choices  $ a=r_1=0.50,\, 0.62$ and $ 0.75 \, \mu$m and 
explore  the angular  behavior of the  magnetic field at the external interface. As  shown in the   Figs. \ref{total5_theta}, \ref{total62_theta} and \ref{total75_theta} the main feature displayed by  the magnetic field is an increase in the angular anisotropy as $r_1$ increases, coupled to a corresponding  increase in its magnitude around  $\theta=\pi/2$. In fact, for $\theta=\pi/2$ we obtain $|\mathbf{B}|\approx 0.20, \, 0.35 $ and $0.60 \, $G, respectively. These results are summarized in the  table \ref{table_B}. Such magnetic field 
strengths can be detected by state-of-the-art diamond magnetometers based on nitrogen-vacancy
(NV) center magnetometry, whose sensitivity can be as high as
$10^{-2} \, {\rm G \, Hz}^{-1/2}$ \cite{36}.  

An alternative possibility to measure the response of the magnetoelectric effect is by  scanning
SQUID (superconducting quantum interference device) magnetometry, which provides very  sensitive detectors  making possible to  measure a change in the  magnetic flux
corresponding to a tiny fraction of one flux quantum $\Phi_0 \,\, (2.2 \times 10^{-8}\, {\rm G cm}^2)$,  typically $ 10^{-6} \, \times  \Phi_0 \,\,  {\rm  Hz}^{-1/2}$ in
today's devices \cite{SQUID}. In general terms, these
devices measure the magnetic flux through a pickup loop. In the Appendix \ref{SQUID}  we have calculated  the magnetic flux over a flat circular loop of radius $R$ located perpendicularly to the $z$ axis   at a distance $\zeta$ of the center of the semiespherical capacitor, as shown in the Fig. \ref{arreglo_loop}. 
 In the table \ref{table_phi} we present  the magnetic flux  through a loop of radius $R=10\mu $m, as a function of  different values of $a=r_1$ and distances $\zeta$. The external interface of the setup is at $r_2= 1\, \mu$m and the flux is given in units of $10^{-10}\, {\rm G} \, {\rm cm}^2$. 
\begin{table}[htb!]
\setlength{\tabcolsep}{10pt}
\renewcommand{\arraystretch}{1.5}
\begin{tabular}{|l|l|l|l|l|l|}
\hline
$a\,(\mu {\rm m})\,\, \backslash \,\, \zeta(\mu{\rm  m})$ & $2$ & $3$ & $4$ & $5$ & $6$ \\ \hline
\hspace{1cm}$0.50$ & $8.4$ & $7.8$ & $7.1$ & $6.4$ & $5.6$ \\ \hline
\hspace{1cm} $0.62$ & $10.2$ & $9.5$ & $8.7$ & $7.8$ & $6.8$ \\ \hline
\hspace{1cm} $0.75$ & $9.9$ & $9.2$ & $8.4$ & $7.5$ & $6.6$ \\ \hline
\end{tabular}
\caption{The magnetic flux, in units of $10^{-10} \, {\rm G cm}^2$, through a circular loop of radius $R=10\mu m$, located at distances $\zeta= 2,\,3,\, 4, \, 5, \, 6 \, \mu$m, for the three configurations previously considered, when the external interface of the setup is at $r_2= 1 \, \mu$m and $a=r_1$.}
\label{table_phi}
\end{table}
On the average we find  fluxes of the order of $10^{-9}\, {\rm G} \, {\rm cm}^2$,  which   comfortably fall  within the measurement capabilities of SQUID devices. 
\acknowledgments
DGV and LFU acknowledge partial support from the project DGAPA-UNAM-IN193319.
\appendix 
\section{The solution of the system of equations}
\label{appendix}
The objective is to find the algebraic expressions that relate the coefficients $A_l^i$, $B_l^i$ , $C_l^i$, $D_l^i$ in
 Eqs. (\ref{gen_elec}
) and  (\ref{gen_mag}
) to the source of the electric excitation given by the potential difference $2V$.
The first step is to rewrite the boundary conditions at the interfaces in terms of the electric and magnetic  potentials.

Let us start with the metallic surface of the capacitor. The particular (hemispheric) form of the electric potential $\Phi$ in $r=a$ is  given by the equation
\begin{equation}
\Phi_1(r=a,\theta)=
\begin{cases}
    +V & \text{if } (0\leq \theta < \pi/2)\\
    -V & \text{if } (\pi/2 < \theta \leq \pi).
\end{cases}
\label{phi_a}
\end{equation}
{Let us observe that the source potential is discontinuous in the equator of the capacitor $r=a, \, \theta=\pi/2$. In this way, the  calculations for observing the fields at points close to this ring, resulting from different choices of our setup, requires a careful limiting process.}

The boundary conditions (\ref{phi_a}) must be implemented in 
\begin{equation}
\Phi_1(r=a, \theta)=\sum_{l=0}^\infty \alpha_l P_l(\cos \theta),
\end{equation}
where $\alpha_l=A_l^1 a^l+B_l^1 a^{-(l+1)}$ needs to be solved in terms of the given potential $V$.  According to Jackson \cite{jackson}, the result is 
\begin{eqnarray}
\alpha_l= A_l^1 a^l + B_l^1 a^{-(l+1)} &=& V_l \quad (\text{odd $l$}), \label{BCPOT}\\
\alpha _l= A_l^1 a^l + B_l^1 a^{-(l+1)} &=& 0 \quad (\text{even $l$}),
\end{eqnarray}
with
\begin{equation}
V_l=V\left(-\frac{1}{2}\right)^{(l-1)/2}\frac{(2l+1)(l-2)!!}{2(\frac{l+1}{2})!}.
\label{V_l}
\end{equation}
This means that  only odd Legendre polynomials appear in the expression of the potentials, that is odd powers of $\cos \theta$, which is in accordance with the symmetry of the problem. Let us recall our notation 
\begin{equation}
\sum_{l}{}^{\prime} \equiv \sum_{l=1,3,...}^\infty.
\end{equation} 
A second condition at the capacitor is that
the normal component of the magnetic field must be zero at $r=a$, since this is the surface of a perfect conductor:
\begin{equation}
\Big(\frac{\partial \Psi_1}{\partial r}\Big)_{r=a}=0. \label{mag_normal}
\end{equation}
At the interfaces $\Sigma_1$ and $\Sigma_2$, the electric  potentials are continuous, i.e.
\begin{eqnarray}\label{nueve}
&& \Big(\Phi_1 = \Phi_2 \equiv \Phi_{1,2}\Big)_{r=r_1}, \quad \Big(\Phi_2 = \Phi_3 \equiv \Phi_{2,3}\Big)_{r=r_2}. \label{diez} 
\end{eqnarray}
When writing two subindices,  $\Theta_{i,j}(R)$ we make explicit that $\Theta_i(R)=\Theta_j(R)$, so it is indifferent if we use the $i$-th or the $j$-th function $\Theta$ for this particular boundary condition. Let us recall that $-\partial \Phi/\partial \theta$ (tangential component of ${\mbf E}$) and $-\partial \Psi/\partial r $  (normal component of ${\mbf B}$) are continuous at the interfaces

Let us now consider the boundary conditions on the derivatives of the potentials which arise from Eqs (\ref{cb}).  The interface at  $r=r_1$ yields
\begin{eqnarray}\label{uno}
&&0=\Bigg(\varepsilon_2 \frac{\partial\Phi_2}{\partial r} - \varepsilon_1\frac{\partial\Phi_1}{\partial r}-  \tilde{ \alpha} \left(\frac{\partial \Psi}{\partial r}\right)_{1,2}\Bigg)_{r=r_1},\\ \label{dos}
&&0=\Bigg(\frac{\partial\Psi_2}{\partial \theta} - \frac{\partial\Psi_1}{\partial \theta} + \tilde{ \alpha} \left(\frac{\partial\Phi}{\partial \theta}\right)_{1,2}\Bigg)_{r=r_1},\\ \label{tres}
&&0=\Big(\frac{\partial\Psi_1}{\partial r} - \frac{\partial\Psi_2}{\partial r}\Big)_{r=r_1},\\ \label{cuatro}
&&0=\Big(\frac{\partial\Phi_1}{\partial \theta} - \frac{\partial\Phi_2}{\partial \theta}\Big)_{r=r_1},
\end{eqnarray} 
while that at  $r=r_2$ produces
\begin{eqnarray} \label{cinco}
&&0=\Bigg(\varepsilon_1 \frac{\partial\Phi_3}{\partial r} - \varepsilon_2\frac{\partial\Phi_2}{\partial r} + \tilde{ \alpha} \left( \frac{\partial \Psi}{\partial r}\right)_{2,3}\Bigg)_{r=r_2},\\ \label{seis}
&&0=\Bigg(\frac{\partial\Psi_3}{\partial \theta} - \frac{\partial\Psi_2}{\partial \theta} - \tilde{ \alpha} \left(\frac{\partial\Phi_{2,3}}{\partial \theta}\right)_{2,3}\Bigg)_{r=r_2},\\ \label{siete}
&&0=\Big(\frac{\partial\Psi_2}{\partial r} - \frac{\partial\Psi_3}{\partial r}\Big)_{r=r_2},\\ \label{ocho}
&&0=\Big(\frac{\partial\Phi_2}{\partial \theta} - \frac{\partial\Phi_3}{\partial \theta}\Big)_{r=r_2}.
\end{eqnarray} 
Finally, both potentials must be zero in infinity if physical solutions are desired. This  condition translates directly into:
\begin{eqnarray}
C_l^3&=&0, \qquad A_l^3=0.
\end{eqnarray}
Before  writing the boundary conditions in terms of the coefficients in Eqs. (\ref{gen_elec}) and (\ref{gen_mag}), a clarification must be made. Take, for example, equation (\ref{cuatro})
after grouping terms
\begin{equation}
\sum_{l}{}^{\prime} \, \Big[\Big(A_l^1 r_1^l+B_l^1 r_1^{-(l+1)}\Big)-\Big(A_l^2 r_1^l+B_l^2 r_1^{-(l+1)}\Big)\Big] \frac{d P_l(\cos \theta)}{d\theta} = 0.
\end{equation}
For simplicity, we define $\gamma_l^i= A_l^i r_1^l+B_l^i r_1^{-(l+1)}$, so that
\begin{equation}
\sum_{l}{}^{\prime} \, \Big(\gamma_l^1 - \gamma_l^2\Big) \frac{d P_l(\cos \theta)}{d\theta} = 0.
\label{pr1}
\end{equation}
Recalling that $dP_l(\cos \theta)/d\theta$ is a combination of the basis $P_l(\cos\theta)$ and  $P_{(l-1)}(\cos\theta)$ it is not immediate whether the above equation  yields  that $(\gamma_l^1 - \gamma_l^2)=0$, or a recurrence relation among these  quantities instead. This question is resolved by proceeding as follows. 
Notice  that
\begin{equation}
\frac{d P_l(\cos \theta)}{d \theta}=\frac{d P_l(\cos \theta)}{d \cos \theta} \frac{d \cos \theta}{d\theta}=\frac{d P_l(\cos \theta)}{d \cos \theta} (-sin(\theta)) .
\end{equation}
So, by multiplying the expression (\ref{pr1}) by ${d P_m(\cos \theta)}/{d \theta}$ and integrating from -1 to 1 we obtain
\begin{equation}
\sum_{l}{}^{\prime} \, \Big(\gamma_l^1 - \gamma_l^2\Big) \int_{-1}^1 (1-x^2) \frac{d P_l(x)}{dx} \frac{d P_m(x)}{dx} dx=0,
\label{gammas}
\end{equation}
with $x=\cos \theta$.
Integrating by parts and using the orthogonality relations of the Legendre polynomials,
\begin{equation}
\int_{-1}^{1} P_l(x)P_m(x)dx = \frac{2}{2l+1}\delta_{lm},
\end{equation}
in addition to the property
\begin{equation}
\frac{d \left[ (1-x^2)\frac{dP_l(x)}{dx}\right]}{dx}=-l(l+1)P_n(x),
\end{equation}
we get to
\begin{eqnarray}
\int_{-1}^{1}(1-x^2) \frac{dP_l(x)}{dx}\frac{dP_m(x)}{dx} dx &=& (1-x^2)\frac{dP_l(x)}{dx} P_m(x) \biggr |_{-1}^{1}-\int_{-1}^{1}-l(l+1)P_l(x)P_m(x)dx \nonumber \\
&=& \frac{2l(l+1)}{2l+1}\delta_{lm}.
\end{eqnarray}
Consequently, the expression (\ref{gammas}) reduces to 
\begin{equation}
\Big(\gamma_l^1 - \gamma_l^2\Big) \frac{2l(l+1)}{2l+1}=0,
\end{equation}
yielding finally 
\begin{equation}
\gamma_l^1 = \gamma_l^2
\end{equation}
for every $l >0$. The case $l=0$ is not a problem since  the only nonzero contributions to the coefficients arise from odd values of $l$, as a consequence of the boundary conditions in the capacitor. In terms of $A_l^i$ and $B_l^i$, the previous relation is
\begin{equation}
A_l^1 r_1^l+B_l^1 r_1^{-(l+1)}=A_l^2 r_1^l+B_l^2 r_1^{-(l+1)},
\label{1a}
\end{equation}
which is the same restriction imposed by the condition $\Phi_1(r_1)=\Phi_2(r_1)$.  Similarly, for the exterior interface $r=r_2$ we obtain
\begin{equation}
A_l^2 r_2^l+B_l^2 r_2^{-(l+1)}=B_l^3 r_2^{-(l+1)}.
\label{2a}
\end{equation}
Then we  conclude that any condition that involves derivatives of the potentials with respect to $\theta$ can be translated into a condition on the potentials themselves. Thus, Eqs. (\ref{cuatro}) and (\ref{ocho}) give rise to the same conditions  as Eq. (\ref{nueve})  .
In the same way, the equations (\ref{dos}) and (\ref{seis})
provide the following boundary conditions 
\begin{eqnarray} \label{disc1}
\Psi_2(r_1)-\Psi_1(r_1)&=&-\tilde{ \alpha} \Phi_{1,2}(r_1),\\ \label{disc2}
\Psi_3(r_2)-\Psi_2(r_2)&=&\tilde{ \alpha} \Phi_{2,3}(r_2),
\end{eqnarray}
whose explicit expressions are  
\begin{eqnarray}
\Big(C_l^2 r_1^l + D_l^2 r_1^{-(l+1)}\Big)-\Big(C_l^1 r_1^l + D_l^1 r_1^{-(l+1)}\Big)&=&-\tilde{ \alpha}\Big(A_l^1 r_1^l + B_l^1 r_1^{-(l+1)}\Big),\\ \label{3a}
D_l^3 r_2^{-(l+1)}-\Big(C_l^2 r_2^l + D_l^2 r_2^{-(l+1)}\Big)&=&\tilde{ \alpha}B_l^3 r_2^{-(l+1)}.
\end{eqnarray}
Having dealt with  all the  relationships involving tangential derivatives, it is necessary to focus our attention on the conditions associated with radial derivatives. Notice that from (\ref{gen_elec}) and (\ref{gen_mag}), we obtain
\begin{eqnarray}
\frac{\partial \Phi_i}{\partial r}&=&\sum_{l}{}^{\prime} \, \Big(l A_l^i r^{l-1} -(l+1) B_l^i r^{-(l+2)}\Big) P_l(\cos \theta),\label{DRPHI}\\
\frac{\partial \Psi_i}{\partial r}&=&\sum_{l}{}^{\prime} \, \Big(l C_l^i r^{l-1} -(l+1) D_l^i r^{-(l+2)}\Big) P_l(\cos \theta). \label{DRPSI}
\end{eqnarray}
So, the  expressions (\ref{uno}) and (\ref{cinco}) give rise to
{\small
\begin{eqnarray}
\varepsilon_2 \Big(l A_l^2 r_1^{l-1} -(l+1) B_l^2 r_1^{-(l+2)}\Big) - \varepsilon_1 \Big(l A_l^1 r_1^{l-1} -(l+1) B_l^1 r_1^{-(l+2)}\Big) &=& \tilde{ \alpha} \Big(l C_l^2 r_1^{l-1} -(l+1) D_l^2 r_1^{-(l+2)}\Big),\nonumber \\  \label{radialphi2} \\
\varepsilon_1 \Big(-(l+1) B_l^3 r_2^{-(l+2)}\Big) - \varepsilon_2 \Big(l A_l^2 r_2^{l-1} -(l+1) B_l^2 r_2^{-(l+2)}\Big) &=& -\tilde{ \alpha} \Big(-(l+1) D_l^3 r_2^{-(l+2)}\Big).
\label{radialphi21}
\end{eqnarray}
}
On the other hand, from the continuity of the radial derivative of the magnetic potential  (\ref{tres}) and (\ref{siete}) it follows that
\begin{eqnarray}
l C_l^1 r_1^{l-1} -(l+1) D_l^1 r_1^{-(l+2)} &=& l C_l^2 r_1^{l-1} -(l+1) D_l^2 r_1^{-(l+2)},\\ \label{4a}
l C_l^2 r_2^{l-1} -(l+1) D_l^2 r_2^{-(l+2)} &=& -(l+1) D_l^3 r_2^{-(l+2)}.
\end{eqnarray}
The equation (\ref{mag_normal}) indicates that
\begin{equation}
l C_l^1 a^{l-1} = (l+1)D_l^1 a^{-(l+2)}. \label{A34}
\end{equation}
There are twelve coefficients for every odd value of $l$ and twelve equations that relate them linearly, which are
\begin{eqnarray}
&A_l^1 a^l + B_l^1 a^{-(l+1)} = V_l, \\ \label{SE}
&A_l^1 r_1^l+B_l^1 r_1^{-(l+1)}= A_l^2 r_1^l+B_l^2 r_1^{-(l+1)},\\
&A_l^2 r_2^l+B_l^2 r_2^{-(l+1)}= B_l^3 r_2^{-(l+1)},\\
&\varepsilon_2 (l A_l^2 r_1^{l-1} -(l+1) B_l^2 r_1^{-(l+2)}) - \varepsilon_1(l A_l^1 r_1^{l-1} -(l+1) B_l^1 r_1^{-(l+2)})= \nonumber \\
& \hspace{5cm} = \tilde{ \alpha} (l C_l^2 r_1^{l-1} -(l+1) D_l^2 r_1^{-(l+2)}),  \label{CER1}\\ 
&\varepsilon_1(-(l+1) B_l^3 r_2^{-(l+2)}) - \varepsilon_2(l A_l^2 r_2^{l-1} -(l+1) B_l^2 r_2^{-(l+2)})=\nonumber \\
& \hspace{5cm} = -\tilde{ \alpha} (-(l+1) D_l^3 r_2^{-(l+2)}),   \label{CER2} \\
&A_l^3=0,
\end{eqnarray}
\begin{eqnarray}
&(C_l^2 r_1^l + D_l^2 r_1^{-(l+1)})-(C_l^1 r_1^l + D_l^1 r_1^{-(l+1)})=-\tilde{ \alpha}(A_l^1 r_1^l + B_l^1 r_1^{-(l+1)}),\label{SMEE1}\\ 
&(D_l^3 r_2^{-(l+1)})-(C_l^2 r_2^l + D_l^2 r_2^{-(l+1)})=\tilde{ \alpha}(B_l^3 r_2^{-(l+1)}), \label{SMEE2}\\
&l C_l^1 r_1^{l-1} -(l+1) D_l^1 r_1^{-(l+2)} = l C_l^2 r_1^{l-1} -(l+1) D_l^2 r_1^{-(l+2)},\\
&l C_l^2 r_2^{l-1} -(l+1) D_l^2 r_2^{-(l+2)} = -(l+1) D_l^3 r_2^{-(l+2)},\\
&l C_l^1 a^{l-1} = (l+1)D_l^1 a^{-(l+2)},\\
&C_l^3 = 0 \label{MF},
\end{eqnarray}
The coefficients $A^i_l$ and   $B^i_l$ correspond to the electric response of the system and their  source is the potential V as shown in Eq. (\ref{SE}). On the other hand, the magnetic coefficients  
 $C^i_l$ and   $D^i_l$ arise due to the magnetoelectric effect  and their source is $\tilde \alpha V$, as shown in Eqs. 
 (\ref{SMEE1}) and (\ref{SMEE2}). Thus  they should vanish when $\tilde \alpha=0$. Then, it is clear from Eqs. (\ref{CER1}) and (\ref{CER2}) that  $A^i_l$ and   $B^i_l$ will receive additional corrections proportional to ${\tilde \alpha}^2 V$. Due to the complexity of the above equations we choose to expand the  solution in powers of $\tilde \alpha$  only up to the the first order. This means we neglect the second order contribution to $A^i_l$ and   $B^i_l$. Nevertheless,  
in the case of 
a TI, $\tilde{ \alpha}$ is proportional to the fine structure constant, being  small enough to justify the validity of the expansion. According to the above strategy, the zeroth order equations for $A_l^i, B_l^i$  are\begin{eqnarray}
&A_l^1 a^l + B_l^1 a^{-(l+1)} = V_l, \\ \label{SE1}
&A_l^1 r_1^l+B_l^1 r_1^{-(l+1)}= A_l^2 r_1^l+B_l^2 r_1^{-(l+1)},\\
&A_l^2 r_2^l+B_l^2 r_2^{-(l+1)}= B_l^3 r_2^{-(l+1)},\\
&\varepsilon (l A_l^2 r_1^{l-1} -(l+1) B_l^2 r_1^{-(l+2)}) - (l A_l^1 r_1^{l-1} -(l+1) B_l^1 r_1^{-(l+2)})=0   \label{CER11}\\ 
&(-(l+1) B_l^3 r_2^{-(l+2)}) - \varepsilon(l A_l^2 r_2^{l-1} -(l+1) B_l^2 r_2^{-(l+2)}) =0\label{CER21} \\
&A_l^3=0, \label{SE11}
\end{eqnarray} 
 where we have introduced the relative permittivity  $\varepsilon=\varepsilon_2/\varepsilon_1$.
 The first order equations in $\tilde \alpha$, for 
 $C_l^i, D_l^i$,
 are
 \begin{eqnarray} 
 &(C_l^2 r_1^l + D_l^2 r_1^{-(l+1)})-(C_l^1 r_1^l + D_l^1 r_1^{-(l+1)})=-\tilde{ \alpha}(A_l^1 r_1^l + B_l^1 r_1^{-(l+1)}),\label{SMEE11}\\ 
&(D_l^3 r_2^{-(l+1)})-(C_l^2 r_2^l + D_l^2 r_2^{-(l+1)})=\tilde{ \alpha}(B_l^3 r_2^{-(l+1)}), \label{SMEE21}\\
&l C_l^1 r_1^{l-1} -(l+1) D_l^1 r_1^{-(l+2)} = l C_l^2 r_1^{l-1} -(l+1) D_l^2 r_1^{-(l+2)},\\
&l C_l^2 r_2^{l-1} -(l+1) D_l^2 r_2^{-(l+2)} = -(l+1) D_l^3 r_2^{-(l+2)},\\
&l C_l^1 a^{l-1} = (l+1)D_l^1 a^{-(l+2)},\label{SMEE111} \\
&C_l^3 = 0.
\end{eqnarray}
The sources of the  linear  equations  (\ref{SMEE11}) and (\ref{SMEE21}) are calculated with the zeroth-order values for $A_l^i, B_l^i$ obtained from the previous set of equations (\ref{SE1}-\ref{SE11}).
Under the above conditions we find 
\begin{eqnarray}
A_l^1 &=& \lambda (l+1)(\varepsilon -1)(l \varepsilon +l+1) \left(r_1^{2 l+1}-r_2^{2 l+1}\right), \label{A51} \\
A_l^2 &=& \lambda (l+1) (2 l+1) (\varepsilon -1) r_1^{2 l+1},\\
B_l^1 &=& -\lambda r_1^{2 l+1} \Big(l (l+1) (\varepsilon -1)^2 r_1^{2 l+1}-r_2^{2 l+1} (l \varepsilon +l+1) (l \varepsilon +l+\varepsilon )\Big),\\
B_l^2 &=& \lambda (2 l+1) r_1^{2 l+1} r_2^{2 l+1} (l \varepsilon +l+1),\\
B_l^3 &=& \lambda (2 l+1)^2 \varepsilon r_1^{2 l+1} r_2^{2 l+1},\\
C_l^1 &=& -\lambda \tilde{ \alpha}  (l+1) (l \varepsilon +l+1) \Big(r_1^{2 l+1}-r_2^{2 l+1}\Big),\\
C_l^2 &=& -\lambda \tilde{ \alpha}  (l+1) (2 l+1) \varepsilon r_1^{2 l+1},\\
D_l^1 &=& -\lambda \tilde{ \alpha}  a^{2 l+1} l (l \varepsilon +l+1) \Big(r_1^{2 l+1}-r_2^{2 l+1}\Big),\\
D_l^2 &=& -\lambda \tilde{ \alpha}  l \Bigg(a^{2 l+1} (l \varepsilon +l+1) \Big(r_1^{2 l+1}-r_2^{2 l+1}\Big)+r_1^{2 l+1} r_2^{2 l+1} (l \varepsilon +l+1)+(l+1) (\varepsilon -1) r_1^{4 l+2}\Bigg),\nonumber \\ \\
D_l^3 &=& -\lambda \tilde{ \alpha}  l \Big(r_1^{2 l+1}-r_2^{2 l+1}\Big) \Big(a^{2 l+1} (l \varepsilon +l+1)+(l+1) (\varepsilon -1) r_1^{2 l+1}\Big), \label{A60}
\end{eqnarray}
where we have defined
\begin{eqnarray}
\lambda&\equiv& \lambda (r_1, r_2, a) =\frac{a^{l+1} V_l}{D} \\
D &=&(l+1) (\varepsilon -1) a^{2 l+1} (l \varepsilon +l+1) \Big(r_1^{2 l+1}-r_2^{2 l+1}\Big) + \nonumber \\
&&+r_1^{2 l+1} r_2^{2 l+1} (l \varepsilon +l+1) (l \varepsilon +l+\varepsilon )-l (l+1) (\varepsilon -1)^2 r_1^{4 l+2}. \label{A62} 
\end{eqnarray}
We verify that in the purely electrical  case, when  $\tilde{ \alpha}=0$, the magnetic coefficients vanish, as expected.
\begin{figure}[htb!]
    \centering
    \includegraphics[scale=0.06]{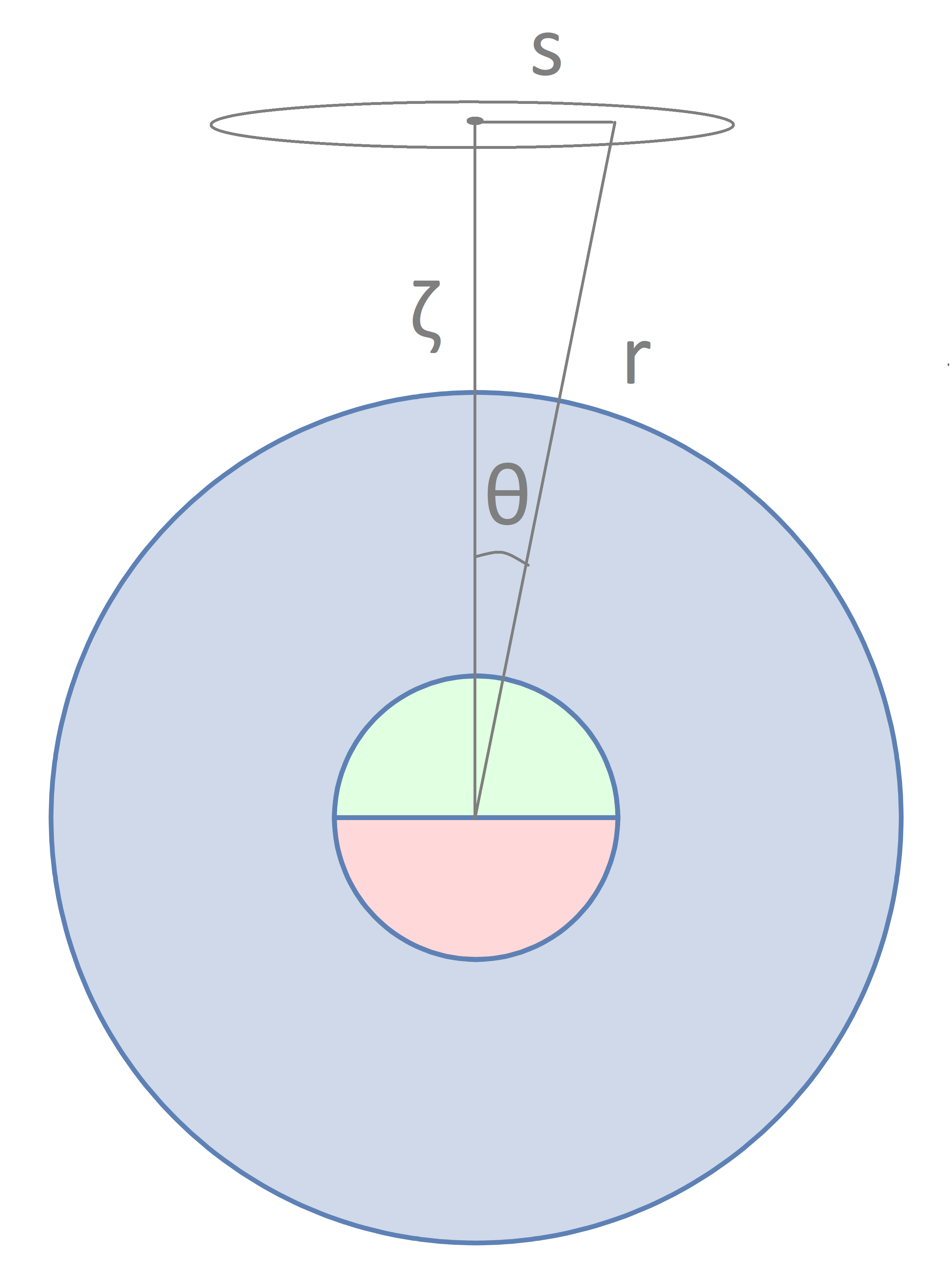}
    \caption{A flat circular loop  positioned perpendicularly to the $z$ axis at a distance $\zeta$ from the center of the semiespherical capacitor.}
    \label{arreglo_loop}
\end{figure}
\section{The magnetic flux over a circular pickup loop}
\label{SQUID}
We calculate the magnetic flux through a flat circular loop of radius $R$ located perpendicularly  to the $z$ axis  which center is at a distance $\zeta$ from the center of the semiespherical capacitor, as shown in Fig. \ref{arreglo_loop}. The distance $s$ form the center of the loop to any point in the region bounded by the loop can be described by means of the polar angle $\theta$ and $\zeta$, according to 
\begin{equation}
s=\zeta \tan \theta,\quad r=\zeta \sec \theta.
\end{equation}
 The position of the points that lie in the interior region of the loop is given by $\mathbf{r}=\zeta \sec \theta \hat{r}$ with $\theta \leq \theta_0= \arctan( {R}/{\zeta}).$
 
In region 3, the magnetic field components  evaluated at the surface bounded by the loop are
\begin{eqnarray}
\label{radial_loop}
B_r(\theta)&\coloneqq B_r(\zeta \sec \theta,\theta)=&\sum_{l}{}^{\prime}\, \left( (l+1)\frac{D_l^3}{(\zeta \sec \theta)^{l+2}} P_l(\cos \theta) \right),\\
\label{tangential_loop}
B_{\theta}(\theta)&\coloneqq B_{\theta}(\zeta \sec \theta,\theta)=& \sum_{l}{}^{\prime}\, \frac{D_l^3}{(\zeta \sec \theta)^{l+2}} \left( \frac{1}{\sin \theta}\left[lP_{l-1}(\cos \theta)-l\cos \theta P_l(\cos \theta) \right] \right).
\end{eqnarray}
On the other hand, the differential area element of the circular loop is given by $d\mathbf{a}=s ds d \phi \hat{k}$,
where $\hat{k}$ is the unit vector in the z-direction. The relation $s=\zeta \tan \theta$ allows us to express the above equation in a different and, although tangled, more useful fashion
\begin{equation}
d\mathbf{a}=\zeta^2 \tan \theta \sec^2 \theta d\theta d\phi \hat{k}.
\end{equation}
Recalling that in spherical coordinates $
\hat{k}=\cos \theta \hat{r}-\sin \theta \hat{\theta}$, 
we obtain 
\begin{equation}
\mathbf{B}\cdot d\mathbf{a}=(B_r(\theta)\cos \theta-B_{\theta}(\theta)\sin\theta)(\zeta ^2 \tan \theta \sec^2\theta d\theta d\phi), 
\label{dot_product}
\end{equation}
with
\begin{equation}
B_r(\theta) \cos \theta -B_\theta(\theta) \sin \theta=\sum_{l}{}^{\prime}\, \frac{D_l^3}{(\zeta \sec \theta)^{l+2}}\left[(2l+1)P_l(\cos \theta)\cos \theta-lP_{l-1}(\cos \theta)\right],
\end{equation}
according to Eqs. (\ref{radial_loop}) and (\ref{tangential_loop}).
The magnetic flux  $
\Phi=\int_{S} \mathbf{B}\cdot d\mathbf{a}
$, with $S$ being  the entire surface bounded by the loop is  
\begin{eqnarray}
\Phi&=&2\pi  \int_0^{\theta_0} \left(\sum_{l=0}^{\infty} \frac{D_l^3}{(\zeta \sec \theta)^l}\left[(2l+1)P_l(\cos \theta)\sin \theta-lP_{l-1}(\cos \theta) \tan \theta \right]\right) d\theta.
\label{FLUX}
\end{eqnarray}
Again we restrict ourselves to the case when the interior interface of the TI touches the plates of the capacitor, i. e  $a=r_1$. Then, the coefficients    $D_l^3$  given by the Eq. 
(\ref{COEFF_CASE_C}). The evaluation of the flux (\ref{FLUX}) proceeds by numerical integration and the results are presented in the table \ref{table_phi}.


\end{document}